 \documentclass[final,5p,times,twocolumn]{elsarticle}

\usepackage{amssymb}
\usepackage{amsmath}

\usepackage{graphicx}	
\usepackage{hyperref}

\DeclareMathAlphabet{\mathpzc}{OT1}{pzc}{m}{it}
\newcommand\beq{\begin{equation}}
\newcommand\eeq{\end{equation}}
\newcommand\bea{\begin{eqnarray}}
\newcommand\eea{\end{eqnarray}}
 
\newcommand\nn{\nonumber\\}

\begin{document}

\begin{frontmatter}


\title{Cosmic magnification in beyond-Horndeski gravity}
\author[a]{Didam G. A. Duniya}
\ead{duniyaa@biust.ac.bw}
\affiliation[a]{organization={Department of Physics \& Astronomy, Botswana International University of Science and Technology}, city={Palapye}, country={Botswana}}

\author[b]{Bishop Mongwane}
\ead{bishop.mongwane@uct.ac.za}
\affiliation[b]{organization={Department of Mathematics \& Applied Mathematics, University of Cape Town}, city={Cape Town 7701}, country={South Africa}}



\begin{abstract}
Cosmic magnification is able to probe the geometry of large-scale structure on cosmological scales, thereby providing another
window for probing theories of the late-time cosmic acceleration. It holds the potential to reveal new information on the nature of dark energy and modified gravity. By using the angular power spectrum, we investigated cosmic magnification beyond weak lensing (incorporating all known relativistic corrections) in beyond-Horndeski gravity---with both constant phenomenology ($\alpha_H$, $\alpha_B$, $\alpha_T$, and $\alpha_K$ being constants) and dynamic phenomenology ($\alpha_H$, $\alpha_B$, $\alpha_T$, and $\alpha_K$ being time-dependent), respectively. For both phenomenologies our results show that the total relativistic signal surpasses cosmic variance (considering an SKA2-like sky coverage) in the magnification angular power spectrum at low redshifts ($z \,{\lesssim}\, 0.5$), hence cosmic-variance reduction methods like multi-tracer analysis will not be needed for surveys at the given $z$. For the individual relativistic signals, we found that the Doppler magnification signal also surpasses cosmic variance and remains the dominant signal, at low $z$, for both phenomenologies. However, the integrated-Sachs-Wolfe, the time-delay, and the gravitational (potential) magnification signals, respectively, are subdominant to both the Doppler magnification signal and cosmic variance, at the same $z$; hence multi-tracer analysis will be needed to isolate these signals. At high redshifts ($z \,{\gtrsim}\, 3$), the integrated-Sachs-Wolfe, the time-delay, and the gravitational magnification signals, respectively, appear to surpass cosmic variance and dominate over the Doppler magnification signal for constant phenomenology; whereas for dynamic phenomenology, all these signals diminish significantly and are well below cosmic variance, at all $z$ (consistent with recent quintessence analysis). Suggesting that including time-variation in the parameters will be crucial in identifying the true signature of the beyond-Horndeski gravity. Furthermore, we found that relativistic effects enhance the ability of the magnification angular power spectrum to probe the imprint of beyond-Horndeski gravity at low redshifts, given their sensitivity to subtle changes in the beyond-Horndeski gravity parameters.
\end{abstract} 


\end{frontmatter}



\section{Introduction}\label{sec:intro}
The beyond-Horndeski gravity \cite{Gleyzes:2014rba, Duniya:2019mpr, Lombriser:2015cla, Sakstein:2016ggl, Gleyzes:2014qga, Gleyzes:2014dya, Bellini:2014fua, Saltas:2019ius, Creminelli:2019kjy, Saltas:2022ybg} is one of the most extensive modified gravity models \cite{Clifton:2011jh, Ishak:2018his} currently in cosmology. Its description of gravity provides a means for a generalized approach of confronting theoretical ideas (of the late-time cosmic accelerated expansion) with current and future observational survey data. The model appears to combine, in a single description, a broad spectrum of well-known existing theories, including dark energy (DE) models \cite{Bamba:2012cp, Duniya:2016gcf, Duniya:2013eta, Duniya:2015nva, Duniya:2015dpa, Copeland:2006wr, Ye:2024ywg, Wolf:2024stt, Wolf:2025jed, Chudaykin:2024gol, Khoury:2025txd, Mohamed:2025ijc}, the scalar-tensor gravity models and their Horndeski extensions, the $f(R)$ gravity and the Horava-Lifshitz gravity models \cite{Clifton:2011jh, Ishak:2018his, Bamba:2012cp}. This theory provides a unified framework for cosmological perturbations about the well-known Friedmann-Robertson-Walker universe, at linear order. Moreover, rather than probing individual models, the cosmological observables of the beyond-Horndeski gravity are instead probed; from where the implication for various models may be inferred. Thus, in view of upcoming precise, large-volume surveys and, given its extensive theoretical reach, it is vital to probe its imprint in the cosmic magnification \cite{Duniya:2016gcf, Mohamed:2025ijc, Bartelmann:1999yn, Bonvin:2008ni, Gillis:2015caa, Umetsu:2015baa, Jeong:2011as, Bacon:2014uja, Raccanelli:2016avd, Andrianomena:2018aad, Duniya:2022vdi, Duniya:2022xcz}. Moreover, recent works (e.g. \cite{Ye:2024ywg, Wolf:2024stt, Wolf:2025jed, Chudaykin:2024gol, Khoury:2025txd}) have found that interacting dark energy (or modified gravity) models are favoured over the well-known cosmological constant and quintessence: hinting towards non-trivial dark sector. (Note, as will be seen, beyond-Horndeski gravity corresponds to an interacting dark energy scenario in which the total energy-momentum tensor is not conserved.)

Cosmic magnification is one of the important observables in modern cosmology. It will be crucial in interpreting measurements of cosmological distances; thus, providing a good means of probing the geometry of large scale structure. The weak (gravitational) lensing \cite{Bartelmann:1999yn, Bonvin:2008ni, Gillis:2015caa, Umetsu:2015baa} is known as the standard source of cosmic magnification. However, cosmic magnification is only one of the effects (along with cosmic shear, see e.g.~\cite{Bonvin:2008ni, Gillis:2015caa, Umetsu:2015baa, Weinberg:2013agg, VanWaerbeke:2009fb, Duncan:2013haa}) of weak lensing. Moreover, weak lensing is not the only source of cosmic magnification, others include \cite{Mohamed:2025ijc, Duniya:2022xcz}: (1) Doppler effect, which is sourced by the line-of-sight relative velocity between the source and the observer, (2) integrated-Sachs-Wolfe (ISW) effect, which is sourced by the integral of the time-rate of change of the gravitational potentials, (3) (cosmological) time delay, which is an integrated effect of the gravitational potentials, and (4) source-observer (gravitational) potential-well effect. Cosmic magnification manifests in large scale structure in various forms, e.g. a source moving towards an observer will appear to experience a boost in flux---this is Doppler magnification \cite{Bacon:2014uja, Raccanelli:2016avd, Andrianomena:2018aad}. Cosmic magnification owing to time-delay effects manifests by broadening of the observed flux. The potential well between the source and the observer can also cause magnification if it is deep enough; particularly, when the observer is at the bottom of the potential well and the source is at the top, e.g. signals from sources with sufficiently lower masses relative to our galaxy (the Milky Way) will appear magnified upon reaching the earth (assuming they are close enough, and other effects are negligible). These last four effects are commonly referred to as ``relativistic effects'' (e.g.~\cite{Duniya:2019mpr, Duniya:2016gcf, Duniya:2013eta, Duniya:2015nva, Duniya:2015dpa, Jeong:2011as, Bacon:2014uja, Raccanelli:2016avd, Andrianomena:2018aad, Duniya:2016ibg, Bonvin:2011bg, Bonvin:2014owa}).

By probing the geometry of large scale structure, cosmic magnification possesses the potential to provide new information on the nature of DE and modified gravity, on the given scales (and redshifts). Moreover, upcoming surveys of large scale structure, which are able to put cosmic magnification to the test, will extend to very large cosmic scales up to near and beyond the Hubble horizon; reaching high redshifts: on these cosmological scales and redshifts, most relativistic effects become significant. Hence, relativistic effects will be key to revealing the strength of cosmic magnification as a cosmological probe, and to understanding the nature of modified gravity (and DE). Thus, in order to realise the full potential of cosmic magnification, we need to correct for the relativistic effects in the observed overdensity of magnified sources (``magnification overdensity,'' henceforth).

In this paper, we study cosmic magnification in the beyond-Horndeski gravity, by probing the angular power spectrum of the observed, relativistic magnification overdensity, on large scales. The goal is not to fit the beyond-Horndeski model to the data (hence warranting quantitative estimations) but to seek a theoretical understanding of cosmic magnification beyond lensing under beyond-Horndeski gravity, and the imprint of the beyond-Horndeski gravity, in the angular power spectrum. This will be crucial for the analysis of cosmological surveys that measure angular sizes of the sources, e.g. upcoming spectroscopic surveys by SKA phase 2 (SKA2) (e.g.~\cite{Maartens:2015mra, Weltman:2018zrl}), Euclid (e.g.~\cite{EUCLID:2011zbd, Amendola:2016saw}), and LSST (e.g.~\cite{LSSTScience:2009jmu, LSSTDarkEnergyScience:2012kar}). It should be pointed out that lensing magnification and total cosmic magnification are commonly interchanged in the literature albeit confusingly; however, in this work we refer to lensing magnification in the true definition (pertaining to the bending of light) and refer to (total) cosmic magnification to include the lensing effect, Doppler effect, and others. The structure of the paper is as follows. In \S\ref{sec:UDE} we discuss the Universe with beyond-Horndeski gravity. In \S\ref{sec:D_mag} we outline the observed (relativistic) magnification overdensity, and in \S\ref{sec:Cls_mag} we analyse the associated angular power spectrum. We conclude in \S\ref{sec:Concl}.


\section{Beyond-Horndeski Gravity}\label{sec:UDE}
In this work, we consider the beyond-Horndeski gravity in the form of a Unified Dark Energy (UDE) \cite{Gleyzes:2014rba, Duniya:2019mpr, Lombriser:2015cla}. Henceforth, we assume a late-time universe dominated by matter (dark plus baryonic) and UDE.

\subsection{The background cosmological equations}
Given the background energy density $\bar{\rho}_A$ and the background pressure $\bar{p}_A$ of matter ($A \,{=}\, m$) and of UDE ($A \,{=}\, x$), the Friedmann equation is given by 
\bea\label{Friedmann}
{\cal H}^2 &=& \dfrac{8\pi G_{\rm eff} a^2}{3}\left(\bar{\rho}_m +\bar{\rho}_x\right),\nn
&=& {\cal H}^2\left(\Omega_m + \Omega_x\right), 
\eea
where $8\pi{G}_{\rm eff} \,{\equiv}\, 1/M^2$, with $M$ being an effective mass; ${\cal H} \,{=}\, a'/a$ is the comoving Hubble parameter, $a \,{=}\, a(\eta)$ is the cosmic scale factor, a prime denotes derivative with respect to conformal time $\eta$ and, $\Omega_m$ and $\Omega_x$ are the matter and the UDE density parameters, respectively.

The associated background acceleration equation, is
\beq\label{acceleration}
 2{\cal H}' + {\cal H}^2 = -8\pi{G}_{\rm eff} a^2 \left(\bar{p}_m+\bar{p}_x\right).
\eeq
The conservation of the total energy-momentum tensor leads to the matter and the UDE background energy density evolution equations, given by
\beq\label{rhoDots}
\bar{\rho}'_m + 3{\cal H}(1 + w_m)\bar{\rho}_m = 0,\quad \bar{\rho}'_x + 3{\cal H}\left(1 + w_x\right)\bar{\rho}_x = \alpha_M {\cal H} \dfrac{\bar{\rho}_x}{\Omega_x},
\eeq
where $w_A \,{=}\, \bar{p}_A/\bar{\rho}_A$ are the equation of state parameters, and 
\beq\nonumber
\alpha_M \equiv \dfrac{2M'}{{\cal H} M} = -\dfrac{G_{\rm eff}'}{{\cal H} G_{\rm eff}},
\eeq
being the mass-evolution parameter, which governs the rate of evolution of $M$; with $G_{\rm eff}$ being as given by \eqref{Friedmann}. 

Consequently, the evolution of the UDE equation of state parameter, is given by
\begin{align}\label{dwxdt}
w'_x =&\; -3{\cal H}\left(1+w_x-\dfrac{\alpha_M}{3\Omega_x}\right)\left(c^2_{ax} - w_x\right),\nn 
=&\; -3{\cal H}\left(1+w_{x,\rm eff}\right)\left(c^2_{ax} - w_x\right),
\end{align}
where $w_{x,\rm eff}$ denotes an ``effective'' equation of state parameter for UDE, and $c^2_{ax} \,{=}\, \bar{p}'_x / \bar{\rho}'_x$ is the square of the adiabatic sound speed of UDE.


\subsection{The perturbed cosmological equations}\label{sec:PEqs}
Here we outline the perturbed field and conservation equations in a universe with UDE. We adopt a flat spacetime metric, given by
\beq\label{metric}
ds^2 = a(\eta)^2 \left[-(1+2\Phi)d{\eta}^2 + (1-2\Psi) d\vec{x}^2\right],
\eeq
where $\Phi$ and $\Psi$ are the (gauge-invariant) temporal and spatial metric potentials, respectively \cite{Gleyzes:2014rba, Duniya:2019mpr} (see also \ref{sec:CEqs}). 

The spatial metric potential evolves according to the equation
\beq\label{PsiDot}
\Psi' + {\cal H}\Phi = -4\pi{G}_{\rm eff}a^2 \sum_A \left(\bar{\rho}_A+\bar{p}_A\right)V_A, 
\eeq
where $V_A$ are the (gauge-invariant) velocity potentials, with the UDE velocity potential being given by
\beq\label{V_x}
8\pi{G}_{\rm eff} \left(\bar{\rho}_x+\bar{p}_x\right) [V_x+\Pi] \;\equiv\; - 2\alpha_B {\cal H} a^{-2} \left(\Pi' + {\cal H}\Pi - \Phi\right),
\eeq
where $\Pi$ is a scalar degree of freedom of the Lagrangian (with dimension of per mass). (This should not be confused with anisotropic stress potentials, as widely used in standard-DE literature.) The metric potentials are related by
\beq \label{PsiPhi}
\Psi - \Phi = 8\pi{G}_{\rm eff}a^2 \sum_A \sigma_A,
\eeq
where $\sigma_A$ are the effective comoving anisotropic stress potentials (having dimension of $M^2$), and $\sigma_x$ is 
\beq\label{sigma_x}
8\pi{G}_{\rm eff} \sigma_x \equiv  a^{-2} \Big[ \alpha_M {\cal H}\Pi - \alpha_T \left(\Psi + {\cal H}\Pi\right) -\alpha_H \left(\Pi' + {\cal H}\Pi - \Phi\right) \Big] ,
\eeq
where $\alpha_H$ is the Horndeski parameter, which measures deviation from Horndeski gravity. The temporal metric potential evolves according to \cite{Duniya:2019mpr}
\beq\label{PhiPrime}
\Phi' + (1+\lambda_1) {\cal H}\Phi = \lambda_2 {\cal H}\Psi + \lambda_3 {\cal H}^2\Pi - 4\pi{G}_{\rm eff}a^2 \lambda_4 \bar{\rho}_m V_m, 
\eeq
where henceforth, we assume matter has zero pressure (i.e. $\bar{p}_m  \,{=}\, 0$, and all pressure-related parameters vanish), and 
\begin{align}\nonumber
\lambda_1 \equiv &\; \alpha_T + \alpha_H(\gamma_5 -\gamma_4) + \alpha_B(1+\alpha_T -\alpha_H\gamma_4) \\ \nonumber
& \; - \dfrac{\alpha'_H}{ {\cal H}\alpha_H} - \beta_1\left(\dfrac{1+\alpha_H}{\alpha_H}\right),\\ \nonumber
\lambda_2 \equiv &\; \dfrac{\alpha'_T}{{\cal H}} - (1+\alpha_T) \dfrac{\alpha'_H}{ {\cal H}\alpha_H} -\alpha_H\gamma_6 - \beta_1\left(\dfrac{1+\alpha_T}{\alpha_H}\right),\\ \nonumber
\lambda_3 \equiv &\; \beta_2 + \beta_1\left(\dfrac{\alpha_M - \alpha_T}{\alpha_H} -1\right),\quad \lambda_4 \equiv 1+\alpha_T - \alpha_H \gamma_4,
\end{align}
where $\alpha_T$ is the tensor speed parameter, which measures the difference between the gravitational waves speed $c_T$ and the speed of light $ c \,{=}\, 1$, with $c^2_T = 1 + \alpha_T$ and, $\alpha_B$ is the kinetic braiding \cite{Gleyzes:2014rba, Duniya:2019mpr, Lombriser:2015cla, Sakstein:2016ggl, Gleyzes:2014qga, Gleyzes:2014dya, Bellini:2014fua}, which measures the kinetic mixing between gravitational and scalar Lagrangian degrees of freedom; with
\begin{align}\nonumber
\beta_1 \equiv &\; \alpha_T + \alpha_B \lambda_4 - \alpha_M - \alpha_H\gamma_1, \\ \nonumber
\beta_2 \equiv &\; (\alpha_M -\alpha_T)\dfrac{\alpha'_H}{\alpha_H {\cal H}} + (\alpha_H -\alpha_M +\alpha_H\gamma_4 - 1)\dfrac{{\cal H}'}{{\cal H}^2} \\ \nonumber
&\; + \dfrac{\alpha'_T -\alpha'_M}{{\cal H}} -\alpha_H\gamma_3 + \lambda_4\left[ 1+\alpha_B - \dfrac{4{\pi}G_{\rm eff}a^2}{{\cal H}^2}\bar{\rho}_m\right],
\end{align}
where $\gamma_1$, $\gamma_3$, $\gamma_4$, $\gamma_5$ and $\gamma_6$ are dimensionless parameters (see \ref{sec:CEqs}). The scalar $\Pi$, given in \eqref{V_x}, evolves by
\beq\label{dpidt}
\Pi' + \left(1 + \dfrac{\alpha_T - \alpha_M}{\alpha_H}\right) {\cal H}\Pi = \left(\dfrac{1+\alpha_H}{\alpha_H}\right) \Phi - \left(\dfrac{1+\alpha_T}{\alpha_H}\right) \Psi .
\eeq
The conservation of the energy-momentum tensor (see \ref{App:T_munu}) leads to the evolution equation for the matter velocity potential $V_m$ and (gauge-invariant) comoving overdensity $\Delta_m$, respectively:
\begin{align}\label{VmEvoln}
V'_m + {\cal H}V_m =&\;  -\Phi ,\\ \label{DmEvoln}
\Delta'_m - \dfrac{9}{2} {\cal H}^2 \Omega_x(1+w_x)\left[V_m - V_x\right] =&\; - \nabla^2V_m ,
\end{align}
where henceforth, we adopt the comoving overdensity $\Delta_A$, which avoids large-scale unphysical anomalies (e.g.~\cite{Bonvin:2011bg, Dent:2008ia}), given by
\beq\label{Delta_A}
\bar{\rho}_A\Delta_A \equiv \delta\rho_A + \bar{\rho}'_A V_A,
\eeq
with $\delta\rho_A$ being the energy density perturbations (see \ref{sec:CEqs} for the expression of $\delta\rho_x$, and $\delta\rho_m$ is prescribed by the energy-momentum tensor in \ref{App:T_munu}).

Similarly, the UDE comoving velocity potential evolves according to the equation:
\beq\label{VxEvoln}
V'_x + {\cal H}V_x = -\Phi - \dfrac{c^2_{sx}\, \Delta_x }{1+w_x} -\dfrac{2\nabla^2\sigma_x}{3(1+w_x)\bar{\rho}_x} - \alpha_M {\cal H} {\cal A},
\eeq
where $\Omega_x(1+w_x) {\cal A} \equiv V_x - \sum_A \Omega_A(1+w_A)V_A$, and $c_{sx}$ is the UDE physical sound speed, given by
\begin{align}
c^2_{sx} =&\; -2\dfrac{(1+\alpha_B)^2}{\alpha_K +6\alpha^2_B} \left\{ 1 + \alpha_T - \dfrac{1 + \alpha_H}{1 + \alpha_B} \Big(2 + \alpha_M - \dfrac{{\cal H}'}{{\cal H}^2} \Big) \right. \nn\nonumber 
& \hspace{1.2cm} \left. - \dfrac{1}{{\cal H}} \Big(\dfrac{1 + \alpha_H}{1 + \alpha_B} \Big)' \right\} - 3\dfrac{(1+\alpha_H)^2}{\alpha_K +6\alpha^2_B} \Omega_m ,
\end{align}
with $\alpha_K$ being the kineticity, which measures the kinetic contribution of the scalar field, and  $\alpha_K \,{+}\, 6\alpha^2_B \,{>}\, 0$.

Although Horndeski gravity ($\alpha_H \,{=}\, 0$), like general relativity, only supports gravitational waves that propagate with luminal speed, beyond-Horndeski gravity ($\alpha_H \,{\neq}\, 0$) is able to support gravitational waves that travel at super-luminal speed ($c_T \,{>}\, 1$) or sub-luminal speed ($c_T \,{<}\, 1$). (See \eqref{alphaT-alphaB}.) Noting that a recent source detection in gravitational and electromagnetic radiation suggested that $\alpha_T \,{\simeq}\, 0$, it is still possible that $\alpha_T$ varies for higher redshift and possibly with frequency (e.g.~\cite{Harry:2022zey, Noller:2018wyv}). Generally, in a Lorentz-invariant system, gravitational waves will propagate at the speed of light \cite{Harry:2022zey}, and this implicitly ensures that the speed of the gravitational waves is independent of scale, time and energy \cite{Noller:2018wyv}. Thus, to admit non-luminal sound speeds, the gravity action (or Lagrangian) will need to be modified to allow time dependence of the cosmological background solution of the scalar degree of freedom---such time dependence will spontaneously break Lorentz invariance and provide a non-trivial medium for gravitational waves to travel through \cite{Harry:2022zey}. These types of modifications that carry time dependence are already enshrined in the beyond-Horndeski gravity. Therefore, the beyond-Horndeski gravity will naturally have $\alpha_T \,{\neq}\, 0$ (since $\alpha_H$ is non-zero); hence, a non-constant gravitational waves speed. If the speed of gravitational waves evolves, then it can be non-luminal. In principle, it can be non-luminal at earlier epochs even if it is tightly constrained to luminal speed at the present cosmic time. Moreover, it is worth noting that the time-dependent, effective mass breaks the total-energy conservation law in the dark sector, as shown by \eqref{rhoDots}, \eqref{VmEvoln}, \eqref{DmEvoln}, \eqref{VxEvoln} and \eqref{DxEvoln}. 

The UDE comoving overdensity evolves according to
\begin{align} \label{DxEvoln}
\Delta'_x - 3 w_x {\cal H}\Delta_x =&\; \dfrac{9}{2} {\cal H}^2 (1+w_x)\sum_A{\Omega_A(1+w_A)\left[V_x - V_A\right]} \nn
& -\; (1+w_x)\nabla^2V_x + \dfrac{2{\cal H}}{\bar{\rho}_x} \nabla^2\sigma_x + \alpha_M {\cal H} {\cal B} ,
\end{align}
where,
\begin{align}\nonumber
\Omega_x {\cal B} \equiv &\; V'_x -\Delta_x + \Big[\dfrac{\alpha'_M}{\alpha_M} -\dfrac{1}{2} (1 + 9w - 2\alpha_M) {\cal H}\Big] V_x \\ \nonumber 
& +\; \sum_A{\Omega_A\Big[\Delta_A -\dfrac{\bar{\rho}'_A}{\bar{\rho}_A}V_A -3{\cal H}(1+w_A)V_A\Big]} .
\end{align}
Notice that \eqref{rhoDots}, \eqref{VxEvoln} and \eqref{DxEvoln} tend to suggest that the given UDE corresponds to an interacting DE scenario in which the total energy-momentum tensor (see \ref{App:T_munu}) is not conserved: $\alpha_M$ breaks the energy-momentum conservation, in the sense that there are extra $\alpha_M$ terms in the UDE conservation equations that do not have counterparts in the matter conservation equations (even with $p_m \,{\neq}\, 0$).

Equations \eqref{acceleration}--\eqref{dwxdt}, \eqref{PsiDot}, \eqref{PhiPrime}--\eqref{DmEvoln}, \eqref{VxEvoln}, and \eqref{DxEvoln} form the complete system of cosmological evolution equations.


\section{The Observed Magnification Overdensity}\label{sec:D_mag}

The observed magnification overdensity \cite{Duniya:2016gcf, Mohamed:2025ijc, Bonvin:2008ni, Jeong:2011as, Duniya:2022vdi, Duniya:2022xcz, Bonvin:2014owa}, measured along the direction ${-}{\bf n}$ at a redshift $z$, is given here by
\beq\label{Delta:obs2}
\Delta^{\rm obs}_{\cal M} ({\bf n},z) \;=\; \Delta^{\rm std}_{\cal M}({\bf n},z) \;+\; \Delta^{\rm rels}_{\cal M}({\bf n},z),
\eeq
where the standard term, is given by
\beq\label{MagStd} 
\Delta^{\rm std}_{\cal M} \equiv -{\cal Q} \int^{r_S}_0{ dr\left(r - r_S\right) \dfrac{r}{r_S} \nabla^2_\perp \left(\Phi + \Psi\right) }, 
\eeq
with $r_S \,{=}\, r(z_S)$ being the radial comoving distance at the source redshift $z_S$ and, ${\cal Q} \,{=}\, {\cal Q}(z)$ being the magnification bias \cite{Mohamed:2025ijc, Duniya:2016ibg, Blain:2001yf, Ziour:2008awn, Schmidt:2009rh, Camera:2013fva, Hildebrandt:2015kcb}, and the relativistic corrections are given by
\beq\label{MagRels}
\Delta^{\rm rels}_{\cal M} = \Delta^{\rm Doppler}_{\cal M} + \Delta^{\rm ISW}_{\cal M} + \Delta^{\rm timedelay}_{\cal M} + \Delta^{\rm potentials}_{\cal M} ,
\eeq
where,
\begin{align}\label{Doppler}
\Delta^{\rm Doppler}_{\cal M} \equiv &\;  -2{\cal Q} \left(1 -\dfrac{1}{r_S {\cal H}}\right)\, ({\bf n}\cdot{\bf V}) , \\ \label{ISW}
\Delta^{\rm ISW}_{\cal M} \equiv &\;   2{\cal Q} \left(\dfrac{1}{r_S {\cal H}} - 1\right) \int^{r_S}_0{dr \left(\Phi' + \Psi' \right) } , \\ \label{timedelay}
\Delta^{\rm timedelay}_{\cal M} \equiv &\;  - \dfrac{2{\cal Q}}{r_S} \int^{r_S}_0{ dr \left(\Phi + \Psi\right)} , \\ \label{potentials}
\Delta^{\rm potentials}_{\cal M} \equiv &\; 2{\cal Q} \left\{\Psi + \left(1 -\dfrac{1}{r_S {\cal H}}\right) \Phi\right\} , 
\end{align}
where ${-}{\bf n}\cdot{\bf V} \,{=}\, -\partial{V}/\partial{r} \equiv V_\parallel$ is the line-of-sight component of the peculiar velocity, with $V$ being the (gauge-invariant) velocity potential.


\section{The Magnification Angular Power Spectrum}\label{sec:Cls_mag}

The total (observed) magnification angular power spectrum observed at $z_S$, is given by
\beq\label{Cl_TT2} 
C_\ell(z_{_S}) = \dfrac{4}{\pi^2} \left(\dfrac{43}{50}\right)^2\int^{r_S}_0 {dk\, k^2 T(k)^2 P_{\Phi_p}(k)\Big|F_\ell(k,z_{_S}) \Big|^2 },
\eeq
where $P_{\Phi_p}$ is the primordial power spectrum, $T(k)$ is the linear transfer function (see e.g.~\cite{Dodelson:2003bk}, for a fitting formula) $k$ is the wavenumber, and we have
\begin{align}\label{f_ell}
F_\ell =&\; \dfrac{{\cal Q}}{r_S} \int^{r_S}_0{dr\, j_\ell(kr) \dfrac{(r-r_S)}{r} \ell(\ell+1) \left(\check{\Phi} + \check{\Psi}\right)(k,r)} \nn
 &+\; 2{\cal Q}\left(\dfrac{1}{r_S {\cal H}} - 1\right) \int^{r_S}_0{dr\, j_\ell(kr) \left(\check{\Phi}'+\check{\Psi}'\right)(k,r)} \nn
 &+\; 2{\cal Q}\left(1 - \dfrac{1}{r_S {\cal H}}\right) \check{V}^\parallel_m \dfrac{\partial}{\partial(kr)} j_\ell(kr_S) \nn
 &+\; 2{\cal Q} j_\ell(kr_S) \left\{ \check{\Psi} +  \left(1 - \dfrac{1}{r_S {\cal H}}\right) \check{\Phi}\right\} \nn
 &-\; \dfrac{2{\cal Q}}{r_S} \int^{r_S}_0{dr\, j_\ell(kr) \left(\check{\Phi} + \check{\Psi}\right)(k,r)} ,
\end{align}
where $j_\ell$ is the spherical Bessel function, and 
\beq\nonumber
\check{\Phi}(k,z) \;{\equiv}\; \Phi(k,z)/\Phi_d(k),
\eeq
similarly for $\check{\Psi}$ and $\check{V}^\parallel_m$; with $\Phi_d$ being the gravitational potential at the photon-matter decoupling epoch ($z \,{=}\, z_d$), given by
\beq\label{Phi_d}
\Phi(k,z_d) = \dfrac{43}{50} \Phi_p(k) T(k) \;{\equiv}\; \Phi_d(k),
\eeq
where $\Phi_p(k)$ is the primordial gravitational potential. Note that we have taken that, given the homogeneity and isotropy on large scales, galaxies flow with the underlying matter (i.e. they have similar velocity profile). We note that the first line in \eqref{f_ell} gives the component for the ``standard'' angular power spectrum $C^{\rm std}_\ell$, which corresponds to \eqref{MagStd}.

To evolve the equations, we consider a scenario which allows the recovery of the well-known $\Lambda$CDM background in early epochs, by setting
\beq\label{w_xeff_val}
w_{x,\rm eff} = -1.
\eeq
A direct consequence of this is that $w_x$ becomes an absolute constant, given \eqref{dwxdt}. A further consequence, is that
\beq\label{aMproptOx}
\alpha_M \propto \Omega_x,
\eeq
where we used the definition of $w_{x,\rm eff}$. Thus, the setup given by \eqref{w_xeff_val} provides a consistent way to connect $\alpha_M$ and $\Omega_x$: the Horndeski parameter becomes directly proportional to the UDE density parameter \eqref{aMproptOx}. Hence, for all numerical computations, we use \eqref{w_xeff_val} and 
\beq\label{params}
\alpha_M = e_M \Omega_x,\quad\; {\cal Q} = 1,
\eeq
where $e_M$ is a constant (with $e_M \,{\leq}\, 0.6$ by imposing $w_{x0} \,{\leq}\, {-0.8}$). It should be further stated that the form of $\alpha_M$ in \eqref{params} does not only allow the recovery of $\Lambda$CDM background in some epochs but also allows the satisfaction of the conservation of total energy and momentum in the given epochs, in both background and perturbations. Particularly, in early epochs $\Omega_x \,{\to}\, 0$, and hence $\alpha_M \,{\to}\, 0$, so that \eqref{rhoDots}, \eqref{VxEvoln} and \eqref{DxEvoln} now obey the conservation of the total energy-momentum tensor. (Note that only $\alpha_M$ is related to energy-momentum conservation.) For emphasis, the form of $\alpha_M$ as given in \eqref{params} is not assumed, but rather arises as a consequence of \eqref{w_xeff_val}. 

For $\alpha_H$, $\alpha_K$, $\alpha_B$, and $\alpha_T$, there is no provision like in \eqref{w_xeff_val} or a fundamental way to determine their form. A common parameterisation is of the form of $\alpha_M$ in \eqref{params}, i.e. set to be proportional to $\Omega_x$ (see e.g. \cite{Bellini:2014fua, Noller:2018wyv}). However, by doing so all distinguishable features of UDE (and hence beyond-Horndeski gravity) vanish at early epochs, where $\Omega_x \,{\to}\, 0$: this removes any possibility of UDE providing or serving as e.g. a non-vanishing early dark energy, which appears to offer a viable solution to the current Hubble and $\sigma_8$ tensions (see e.g. \cite{Poulin:2018cxd, Kamionkowski:2022pkx, Poulin:2023lkg, Niedermann:2023ssr, Brissenden:2023yko, Shen:2024hpx}). In this work, we use the ansatz:
\beq\label{alphas}
\alpha_i = e_i \left(1+\Omega_x\right),
\eeq
where $e_i$ ($i \,{=}\, H,\, K,\, B,\, T$) are absolute constants. Thus, \eqref{alphas} extends previous parameterisation (i.e. $\alpha_i/\Omega_x$ = constant).

\begin{figure*}[!h]\centering
\includegraphics[scale=0.4]{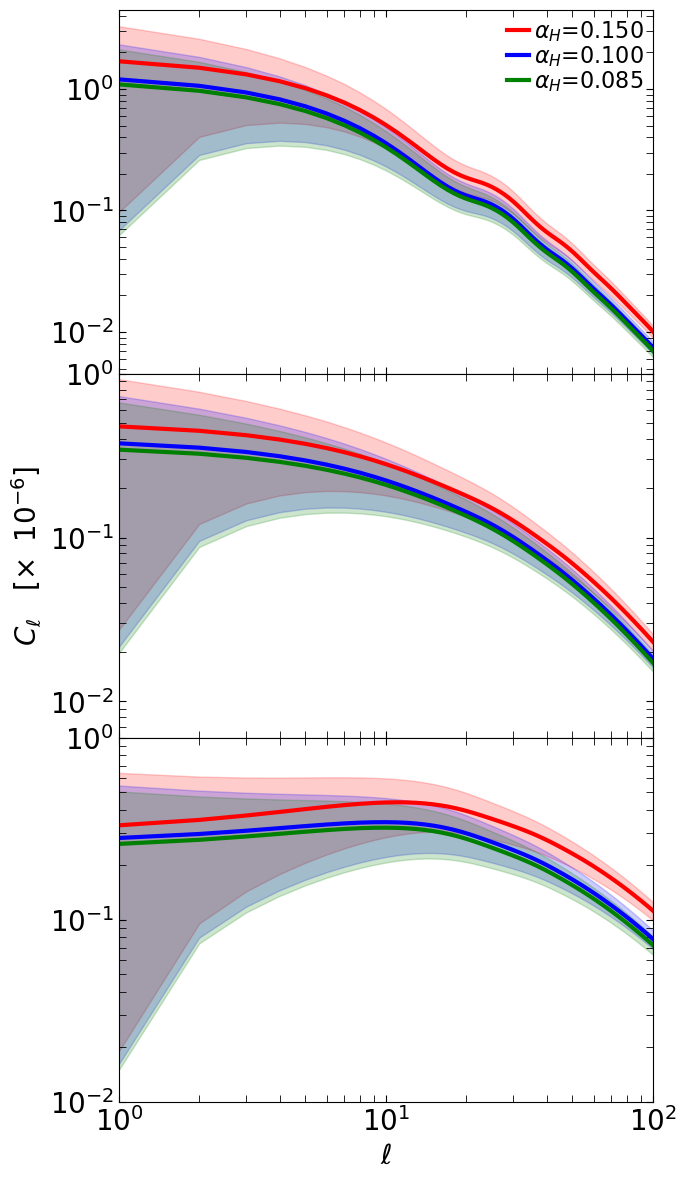} \includegraphics[scale=0.4]{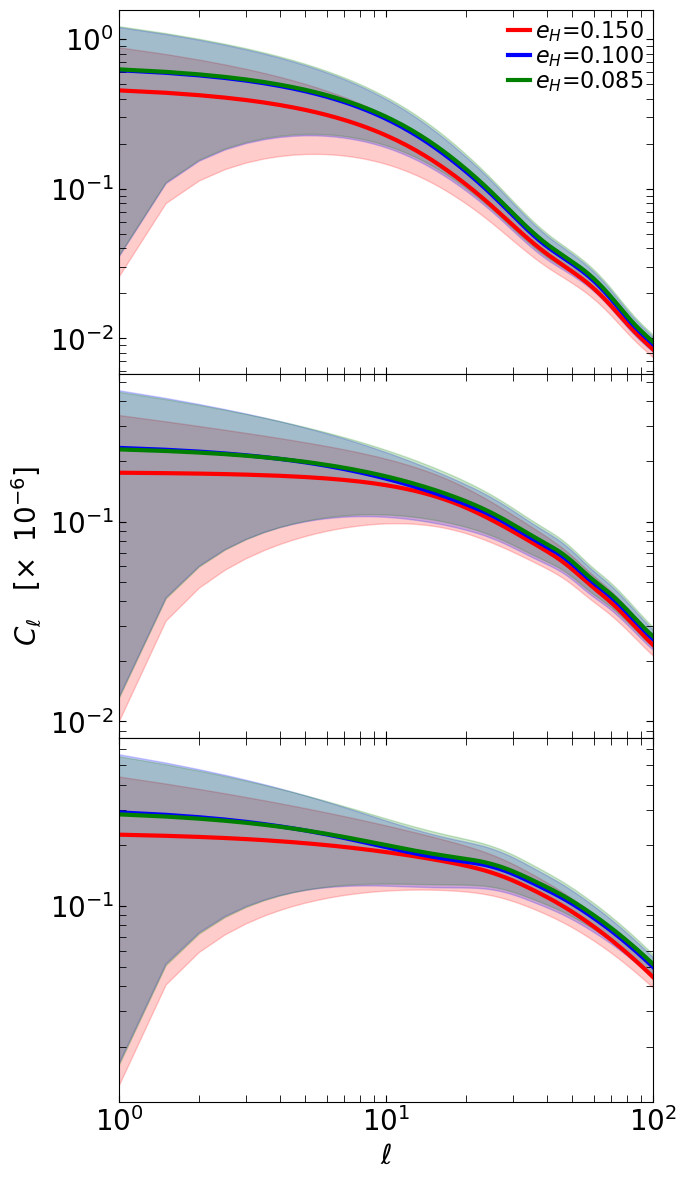} 
\caption{The plots of the total magnification angular power spectrum $C_\ell$ as a function of multipole $\ell$. \emph{Left:} Plots for constant $\alpha_i$ ($i \,{\neq}\, M$) scenario \eqref{case1}, where the Horndeski parameter $\alpha_H \,{=}\, e_H \,{=}\, 0.085,\, 0.1,\, 0.15$ and, $\alpha_B \,{=}\, e_B$ and $\alpha_T \,{=}\, e_T$ are given by \eqref{alphaT-alphaB}. \emph{Right:} Plots for dynamic $\alpha_i$ scenario \eqref{case2}, where $\alpha_H \,{\neq}\, e_H \,{=}\, 0.085,\, 0.1,\, 0.15$ and, $\alpha_B \,{\neq}\, e_B$ and $\alpha_T \,{\neq}\, e_T$, with $\alpha_H$, $\alpha_K$, $\alpha_B$, and $\alpha_T$ being given by \eqref{alphas} and $e_B$ and $e_T$ being given by \eqref{alphaT-alphaB}. For all numerical computations, $\alpha_M$ is dynamic and given by \eqref{params} with $\alpha_M \,{\neq}\, e_M \,{=}\, 0.06$, and $\alpha_K \,{=}\, e_K \,{=}\, 0$. The different panels in both \emph{Left} and \emph{Right} are at the source redshifts $z_S \,{=}\, 0.5$ ({\it top}), $z_S \,{=}\, 1.0$ ({\it middle}), and $z_S \,{=}\, 3.0$ ({\it bottom}), respectively. Shaded regions show the extent of cosmic variance.}\label{fig:totalCls}
\end{figure*}

Evolutions are initialized at the photon-matter decoupling epoch $1\,{+}\,z_d \,{=}\, 10^3 \,{=}\, 1/a(z_d)$, and we use adiabatic initial conditions (e.g.~\cite{Duniya:2019mpr}) for the perturbations. We also use a matter density parameter $\Omega_{m0} \,{=}\, 0.3$ and a Hubble constant $H_0 \,{=}\, 67.8\, {\rm km / s / Mpc}$, from the Planck 2015 results \cite{Planck:2015fie}. We choose the values of $e_M$, $e_H$ and $e_K$ freely, and infer those of $e_B$ and $e_T$ from astrophysical constraints \cite{Sakstein:2016ggl}, given by (see also \cite{Duniya:2019mpr}) 
\beq\label{alphaT-alphaB}
e_B = \left(1-5\dfrac{\Upsilon_2}{\Upsilon_1}\right) e_H,\quad e_T = \dfrac{4e^2_H + (e_H-e_B)\Upsilon_1}{\left(1+e_B\right)\Upsilon_1}, 
\eeq
where $\Upsilon_1$ is the parameter that governs the deviation from Newton's law in astrophysical systems, and $\Upsilon_2$ is the parameter that governs light bending around non-relativistic objects (both parameters are dimensionless constants); with $\Upsilon_1 \,{=}\, {-}0.11^{+0.93}_{-0.67}$ and $\Upsilon_2 \,{=}\, {-}0.22^{+1.22}_{-1.19}$ \cite{Sakstein:2016ggl} (see also~\cite{Saltas:2019ius, Saltas:2022ybg}). Note that although it appears, by observational constraints \eqref{alphaT-alphaB}, as though $c^2_T \,{=}\, 1 \,{+}\, \alpha_T \,{=}\, 1$ or $\alpha_T \,{=}\, 0$ when $\alpha_H \,{=}\, 0$ (Horndeski gravity), this in general is not true, as $\alpha_T \,{\neq}\, 0$ can be achieved with modified Lagrangian terms (see e.g. \cite{Bellini:2014fua, Baker:2017hug}, for details).

Moreover, we choose values of $e_H$ such that we recover the same values of $H_0$ and $\Omega_{m0}$, with $e_M$ and $e_K$ set to values that prohibit superluminal sound speeds, and also allow the recovery of luminal sound speeds at some epochs (see e.g.~\cite{Duniya:2019mpr}). By the choice of the values of $e_H$, the magnification angular power spectra will match on small scales, at today: any deviations that are solely owing to $\alpha_H$ will be isolated on the largest scales; at earlier epochs ($z \,{>}\, 0$) the angular power spectra will separate on all scales. Furthermore, given that the density of astrophysical objects decreases radially outwards from the centre so that $\Upsilon_1 \,{>}\, 0$ (${<}\, 0$) corresponds to weakening (strengthening) gravity \cite{Duniya:2019mpr, Sakstein:2016ggl}, we use $\Upsilon_1 \,{=}\, 0.78$ and adopt a corresponding value for $\Upsilon_2 \,{=}\, {-}0.131$. 

Although from the work by \cite{Creminelli:2019kjy} it appears as though $|\alpha_H| \gtrsim 10^{-20}$ may be ruled out by stability conditions, it should be pointed out that the analysis in the given work is physically limited and not applicable to our analysis, in the sense that it is based on the assumption that there is no matter present (in both background and perturbations), and is also confined to sub-Hubble scales. In essence, the given constriants will only hold in the vicinity of an isolated, compact astrophysical object or a source of gravitational waves (or a gravitational-waves dominated universe). Any attempt to extrapolate these results directly into predictions for large-scale phenomena at distances beyond compact-object scales will be inconsistent and impractical. A proper analysis will have to incorporate matter in the background, with matter-source inhabited large scales, as is evident in our Universe (where matter sources are ubiquitous).

\begin{figure*}[!h]\centering
\includegraphics[scale=0.4]{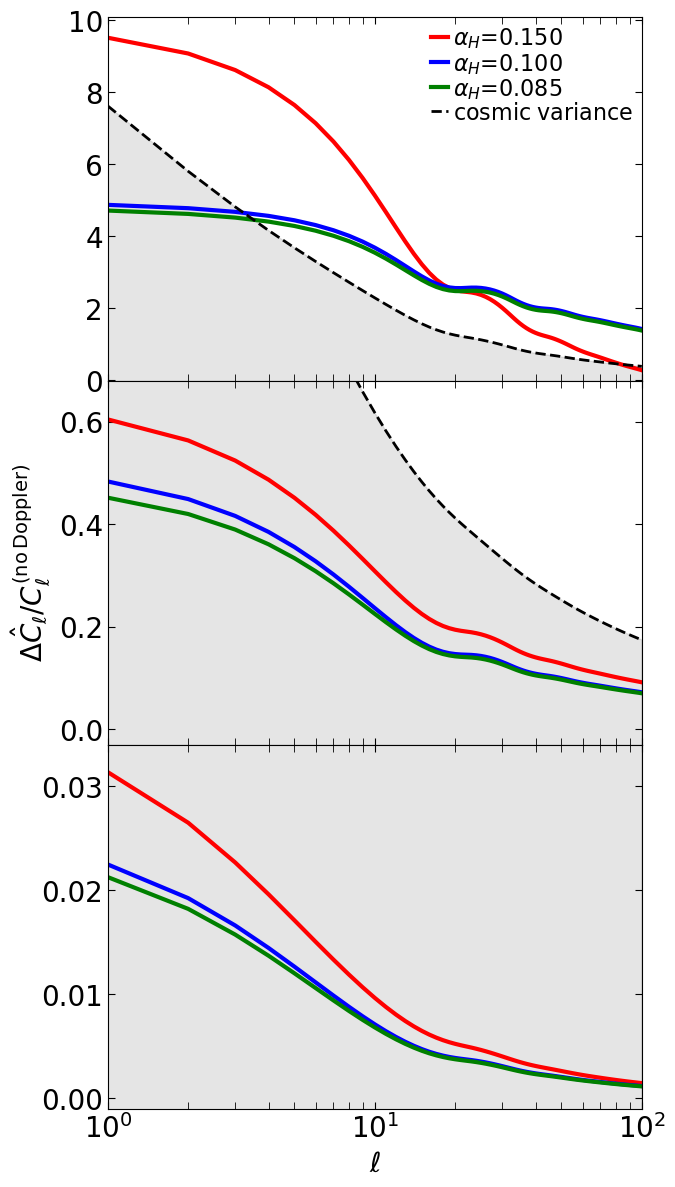} \includegraphics[scale=0.4]{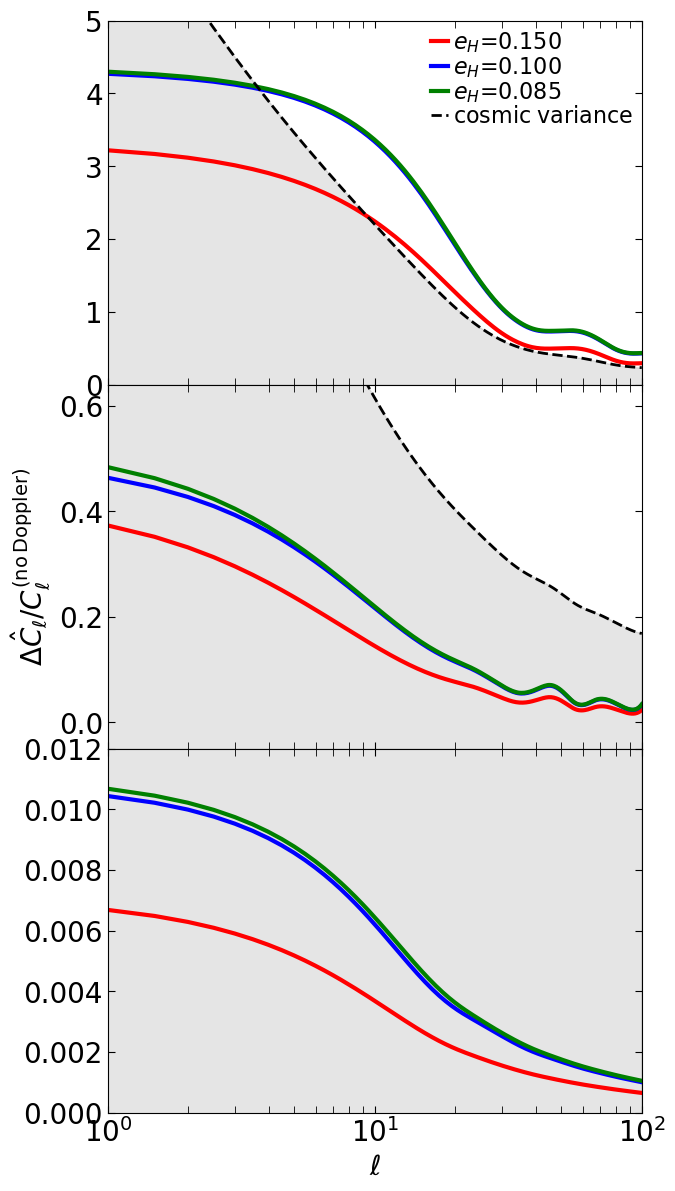} 
\caption{The plots of Doppler signal in the total magnification angular power spectrum $C_\ell$ for the same UDE parameters and scenarios as in Fig.~\ref{fig:totalCls}. \emph{Left:} Plots for constant $\alpha_i$ ($i \neq M$) scenario \eqref{case1}. \emph{Right:} Plots for dynamic $\alpha_i$ scenario \eqref{case2}. Similarly, the \emph{Left} and \emph{Right} panels show the plots at $z_S \,{=}\, 0.5$ ({\it top}), $z_S \,{=}\, 1.0$ ({\it middle}), and $z_S \,{=}\, 3.0$ ({\it bottom}), where $\Delta{\hat C}_\ell \,{=}\, C_\ell \,{-}\, C^{(\rm no\, Doppler)}_\ell$.}\label{fig:DopplerEffect}
\end{figure*}

In view of the goal of this work, we adopt cosmic variance as the source of error in the angular power spectrum. The cosmic variance, is given by (e.g.~\cite{Duniya:2019mpr, Duniya:2022xcz})
\beq\label{sigma}
\sigma_\ell(z) = \sqrt{\dfrac{2}{(2\ell+1)f_{\rm sky}}}\, C_\ell(z),
\eeq
where $f_{\rm sky}$ is the observed sky fraction. Here we adopt the value of an SKA2-like spectroscopic survey, $f_{\rm sky} \,{=}\, 0.75$. Proper quantitative methods and error (or noise) analysis will need to be taken into account, for practical purposes. Moreover, in probing the magnification angular power spectrum \eqref{Cl_TT2} (Figs.~\ref{fig:totalCls}--\ref{fig:potential-SN}), we consider two cases: 
\bea\label{case1}
\mbox{Case~I (Constant $\alpha_i$):} && \alpha_i = e_i,\\ \label{case2}
\mbox{Case~II (Dynamic $\alpha_i$):} && \alpha_i\,/\,(1+\Omega_x) = e_i,
\eea
where $e_i$ are absolute constants ($\alpha_i \,{\neq}\, \alpha_M$); with $e_B$ and $e_T$ being as in \eqref{alphaT-alphaB} and, $\alpha_M/\Omega_x \,{=}\, e_M \,{=}\, 0.06$ and $\alpha_K \,{=}\, e_K \,{=}\, 0$ for all numerical computations.

In Fig.~\ref{fig:totalCls} (left) we consider Case~I \eqref{case1} for the total magnification angular power spectrum \eqref{Cl_TT2}. We give the plots of the total magnification angular power spectrum for the Horndeski parameter values $\alpha_H \,{=}\, e_H \,{=}\, 0.085,\, 0.1,\, 0.15$: at the source redshifts $z_S \,{=}\, 0.5$ (top panel), $z_S \,{=}\, 1$ (middle panel), and $z_S \,{=}\, 3$ (bottom panel). We also indicate the associated extent of cosmic variance \eqref{sigma}. We observe that, on scales $\ell \,{\lesssim}\, 20$, the amplitude of the angular power spectrum decreases as the source redshift increases; whereas, on smaller scales ($\ell \,{>}\, 20$), the amplitude increases as the source redshift increases. Moreover, the angular power spectrum for the different values of the Horndeski parameter gradually diverge on scales $\ell \,{>}\, 20$, as source redshift increases; with the separation becoming more prominent at $z_S \,{\gtrsim}\, 3$, for $\alpha_H \,{>}\, 0.085$. Furthermore, we see that the amplitude of the angular power spectrum increases with higher values of $\alpha_H$, at all values of $z_S$ and $\ell$. This implies that cosmic magnification becomes enhanced (on all scales) as the strength of gravity increases. Thus, for beyond-Horndeski gravity with constant $\alpha_H$, stronger gravity regimes will induce stronger magnification events, while weaker gravity regimes induce (relatively) weaker magnification. Similarly, in Fig.~\ref{fig:totalCls} (right) we consider Case~II \eqref{case2} for the total magnification angular power spectrum, for $\alpha_H \,{\neq}\, e_H \,{=}\, 0.085,\, 0.1,\, 0.15$ at the same $z_S$ values. We see that the overall features of the results for constant $\alpha_H$ (left) and dynamic $\alpha_H$ (right) look similar, except that for dynamic $\alpha_H$ the amplitude of the total magnification angular power spectrum is relatively lower for the same amplitude values of $\alpha_H$, with that for $\alpha_H \,{\neq}\,e_H \,{=}\, 0.15$ particularly becoming subdominant (unlike for $\alpha_H \,{=}\, e_H \,{=}\, 0.15$ where it gives the dominant effect).

Moreover, the shaded regions in the plots of Fig.~\ref{fig:totalCls} (both left and right) show the extent of cosmic variance for the respective chosen values of the Horndeski parameter. As is already widely known, we see that cosmic variance becomes significant with increasing scale (decreasing $\ell$) and increasing source redshift, encompassing the cosmic magnification signal in the angular power spectrum. This implies that the cosmic magnification signal in beyond-Horndeski gravity will ordinarily be overshadowed by cosmic variance in the analysis. Consequently, this signal will (in principle) not be observed on the largest scales, at the given source redshifts. However, by applying multi-tracer analysis (see e.g.~\cite{Abramo:2013awa, Alonso:2015sfa, Fonseca:2015laa, Witzemann:2018cdx, Paul:2022xfx, Karagiannis:2023lsj, Zhao:2023ebp}, for the multi-tracer framework), the effect of cosmic variance can be subdued with large-volume, high-precision future surveys; thereby availing the possibility of detecting the magnification signal. It should be pointed out that actual measurability of any signals will be subject to proper quantitative analysis, which includes several factors e.g. effectiveness and efficiency of the quantitative methods, survey specifications, error modelling, and properties of the sources. (This is outside the goal of this paper, as stated in \S\ref{sec:intro}.)


\subsection{Imprint of UDE and total relativistic effect}
\label{subsec:UDE-imprint}
Here we discuss the UDE imprint in the total (relativistic) magnification angular power spectrum \eqref{Cl_TT2}. The focus is on highlighting the qualitative effect of UDE with respect to relativistic effects, on very large scales. 

\begin{figure*}[!h]\centering
\includegraphics[scale=0.4]{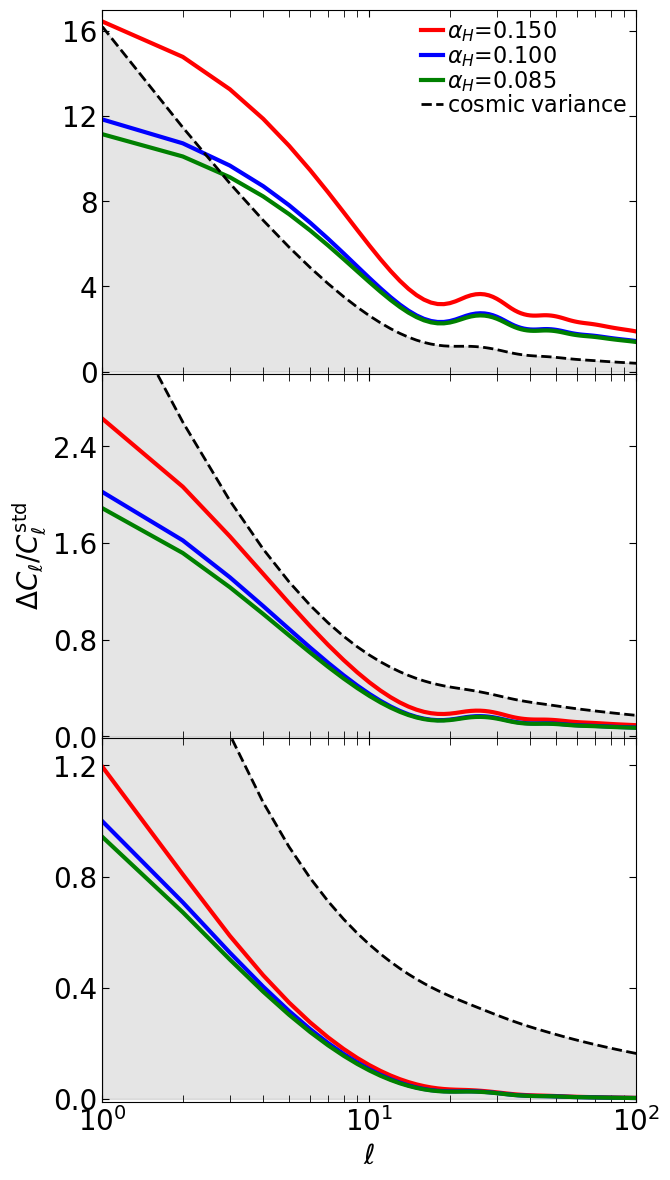} \includegraphics[scale=0.4]{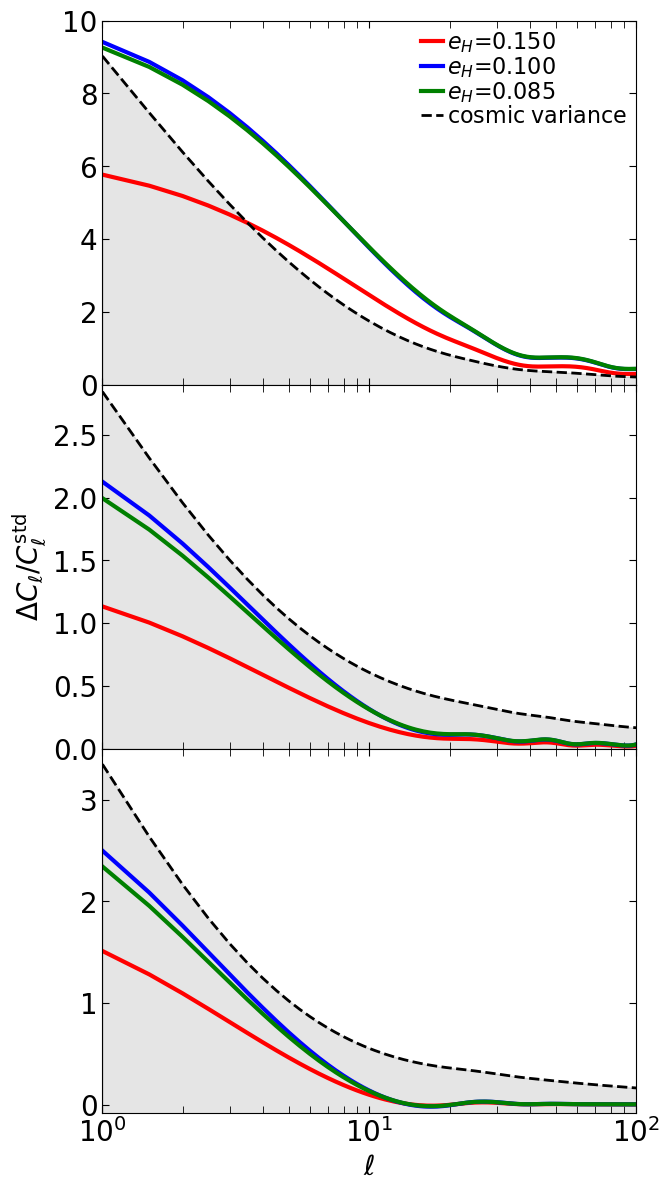}
\caption{The plots of the total relativistic signal in the magnification angular power spectrum $C_\ell$ for the same UDE parameters and scenarios as in Fig.~\ref{fig:totalCls}. \emph{Left:} Plots for constant $\alpha_i$ ($i \neq M$) scenario \eqref{case1}. \emph{Right:} Plots for dynamic $\alpha_i$ scenario \eqref{case2}. The \emph{Left} and \emph{Right} panels show the plots at $z_S \,{=}\, 0.5$ ({\it top}), $z_S \,{=}\, 1.0$ ({\it middle}) and $z_S \,{=}\, 3.0$ ({\it bottom}), where we have $\Delta{C}_\ell \,{=}\, C_\ell - C^{\rm std}_\ell$.}\label{fig:tot2stdClsFracs}
\end{figure*}

In Fig.~\ref{fig:tot2stdClsFracs} we show the plots of the fractional change $\Delta{C}_\ell / C^{\rm std}_\ell$ in the total magnification angular power spectrum $C_\ell$ relative to the standard (weak-lensing) angular power spectrum $C^{\rm std}_\ell$, where $\Delta{C}_\ell \,{=}\, C_\ell \,{-}\, C^{\rm std}_\ell$, for the same scenarios as in Fig.~\ref{fig:totalCls}: Constant $\alpha_H$ \eqref{case1} (left) and dynamic $\alpha_H$ \eqref{case2} (right). Similarly, the plots are at source redshifts $z_S \,{=}\, 0.5$ (top panel), $z_S \,{=}\, 1$ (middle panel), and $z_S \,{=}\, 3$ (bottom panel). The given fractions measure the total effect of relativistic corrections \eqref{Doppler}--\eqref{potentials}, i.e.~the total relativistic signal, in the total magnification angular power spectrum for the given parameters and source redshifts. We also show the reach of cosmic variance (shaded regions) for the total relativistic signal, given by 
\beq\label{sigma_eff}
\sigma^{\rm rels}_\ell = \dfrac{\sqrt{2} }{C^{\rm std}_\ell} \sigma_\ell,
\eeq
where we obtain this relation by using \eqref{sigma} and error propagation theory (see e.g.~\cite{Bevington:2003eta, Wikipedia:2024pou}); with $\sigma_\ell$ being as given by \eqref{sigma} (taking that $C_\ell$ and $C^{\rm std}_\ell$ are uncorrelated). The regions of cosmic variance \eqref{sigma_eff} in Fig.~\ref{fig:tot2stdClsFracs} are done with the least Horndeski parameter, $\alpha_H \,{=}\, 0.085$, at each $z_S$. 

For both constant $\alpha_H$ (left panels) and dynamic $\alpha_H$ (right panels), we see that on moving from high to low source redshift (bottom to top), the amplitude of the total relativistic signal in the total magnification angular power spectrum for each value of the amplitude of the Horndeski parameter increases. This implies that the amount of total relativistic signal in cosmic magnification will be higher at low source redshifts ($z_S \,{<}\, 1$) than at high source redshifts ($z_S \,{\geq}\, 1$), in beyond-Horndeski gravity. Similar results were previously found for interacting DE \cite{Duniya:2016gcf}. Moreover, the total relativistic signal gradually becomes significant with respect to cosmic variance as source redshift decreases: at $z_S \,{\geq}\, 1$, the amplitude of the total relativistic signal is below cosmic variance on all scales; whereas, at $z_S \,{=}\, 0.5$ the amplitude of the total relativistic signal surpasses cosmic variance on scales $\ell \,{\gtrsim}\, 3$. This implies that at low source redshifts ($z_S \,{\leq}\, 0.5$) cosmic-variance reduction in the quantitative analysis of the total relativistic signal in the magnification angular power spectrum will not be necessary; whereas, at high source redshifts ($z_S \,{\geq}\, 1$), multi-tracer methods will be required to beat down cosmic variance. Thus, the data analysis of cosmological surveys that depend on the apparent flux or angular size of sources, such as spectroscopic surveys of SKA2, Euclid, and LSST, in principle can suitably neglect the cosmic-variance effect when analysing the total signal of non-lensing effects in cosmic magnification in beyond-Horndeski gravity, at low redshifts ($z \,{<}\, 0.5$). This underscores the importance of wide-angle, low-redshift surveys. Moreover, we see that the time-dependent phenomenology of dynamic $\alpha_H$ (right panels) will lead to relatively lower amplitudes of the total relativistic signal, at $z_S \,{<}\, 3$; whereas at $z_S \,{\geq}\, 3$ the amplitudes of the total relativistic signal for dynamic $\alpha_H$ become relatively larger. Also, dynamic $\alpha_H$ gives a subdominant amplitude of the total relativistic signal for larger $e_H$ (${\geq}\, 0.15$) at all $z_S$, which is surpassed by cosmic variance on scales $\ell \,{\lesssim}\, 3$. This is the exact converse of the constant $\alpha_H$ scenario (left panels). These results are consistent with those in Fig.~\ref{fig:totalCls}. 

Furthermore, we see that at each source redshift the amplitude of the total relativistic signal increases as the value of the Horndeski parameter increases throughout for constant $\alpha_H$ (left panels) and on the other hand, for dynamic $\alpha_H$ (right panels) the total relativistic signal first increases for an initially growing amplitude of $\alpha_H$ ($e_H \,{<}\, 0.15$) and then reverses (becomes suppressed) at a relatively extreme amplitude ($e_H \,{=}\, 0.15$). This shows that an increasing gravity strength in beyond-Horndeski theory will lead to a boost or enhancement of the total relativistic signal at all times for constant $\alpha_H$ but will lead to an irregular effect---an enhancement for a growing $\alpha_H$ amplitude up to a certain maximum, and a suppression beyond this maximum---for dynamic $\alpha_H$ (for the given choice of $\alpha_i$ parameterisation and parameter values). Moreover, the separation in the total relativistic signal for successive values of the Horndeski parameter increases with decreasing source redshift. In other words, the total relativistic signal for the various values of the Horndeski parameter become relatively better differentiated at lower source redshifts. This shows that the signal of the combined relativistic effects is sensitive to small changes in UDE, and becomes more sensitive at lower redshifts. Thus, relativistic effects will be crucial in identifying the imprint of UDE, and possibly putting constraints on UDE, at late-time epochs ($z_S \,{\leq}\, 0.5$). In general, the results in Fig.~\ref{fig:tot2stdClsFracs} are consistent with those for quintessence by \cite{Mohamed:2025ijc}, where it was found that the amplitude of the total relativistic signal in cosmic magnification will be higher at low source redshifts ($z_S \,{\lesssim}\, 0.5$), and the signal can be detected at the given $z_S$ (subject proper quantitative error analysis).

\subsection{Individual relativistic effects}
\label{subsec:rels-effects}
In this section, we look at the contribution of the individual relativistic corrections \eqref{Doppler}--\eqref{potentials}, in the total magnification angular power spectrum.

\begin{figure*}[!h]\centering
\includegraphics[scale=0.4]{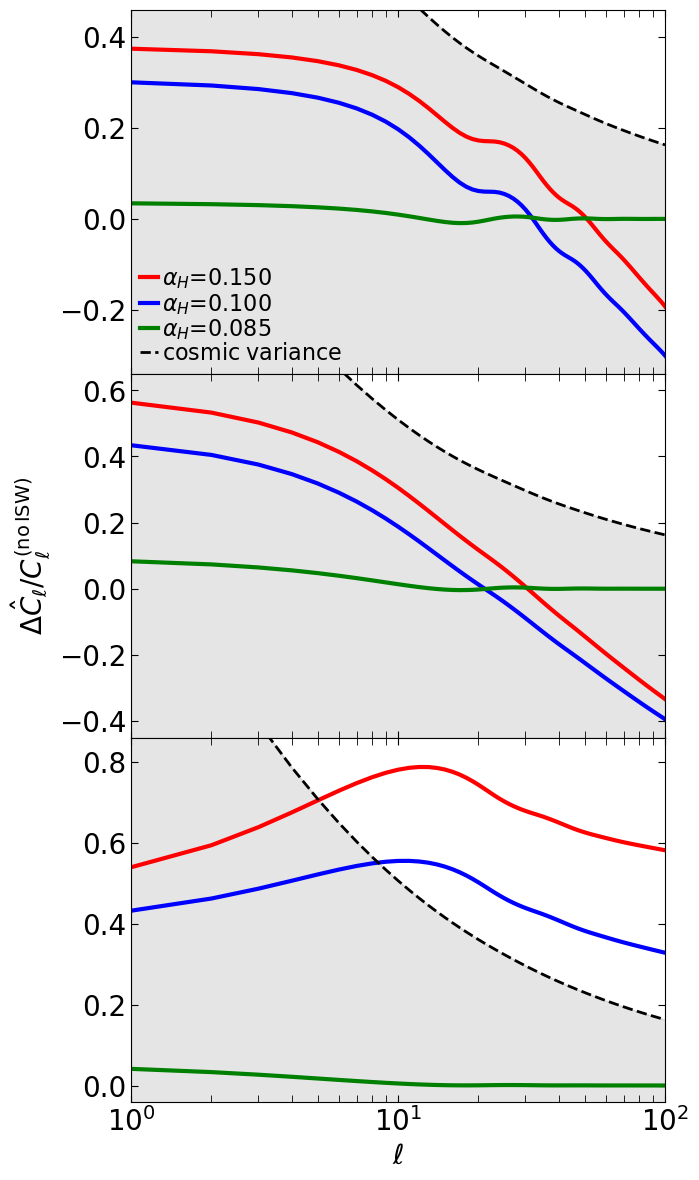} \includegraphics[scale=0.4]{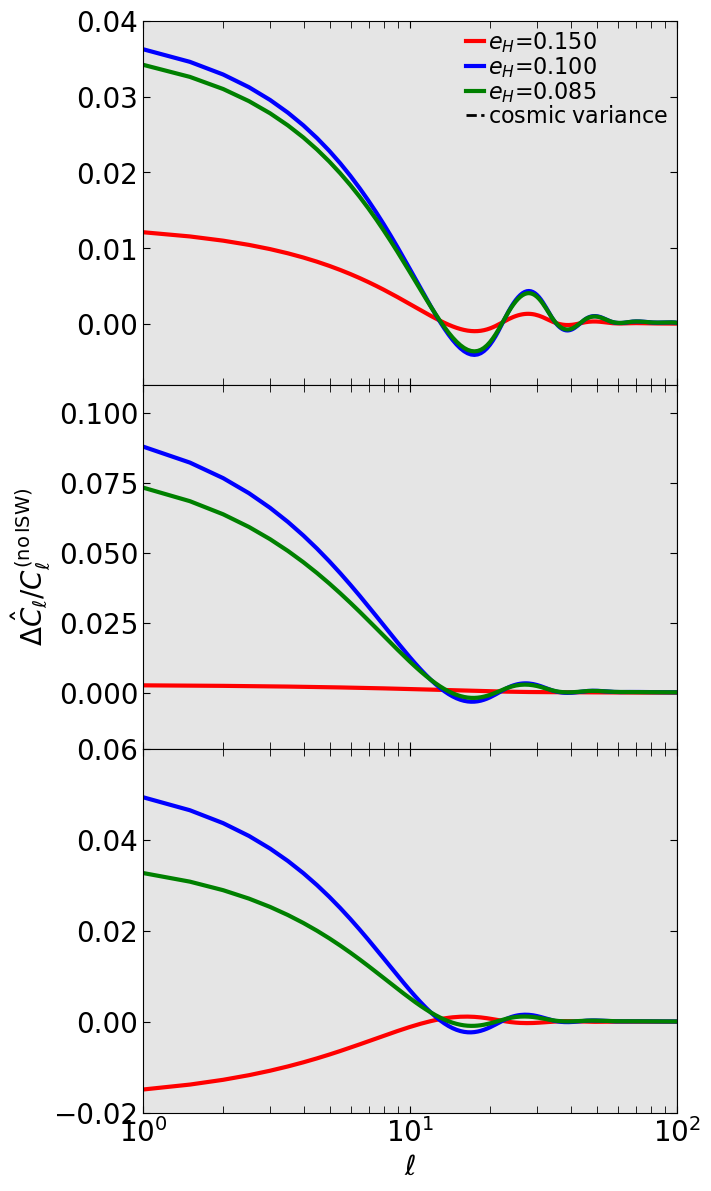} 
\caption{The plots of ISW signal in the total magnification angular power spectrum $C_\ell$ for the same UDE parameters and scenarios as in Fig.~\ref{fig:totalCls}. \emph{Left:} Plots for constant $\alpha_i$ ($i \neq M$) scenario \eqref{case1}. \emph{Right:} Plots for dynamic $\alpha_i$, scenario \eqref{case2}. The plots in the \emph{Left} and \emph{Right} panels at $z_S \,{=}\, 0.5$ ({\it top}), $z_S \,{=}\, 1.0$ ({\it middle}), and $z_S \,{=}\, 3.0$ ({\it bottom}), where we have $\Delta{\hat C}_\ell \,{=}\, C_\ell \,{-}\, C^{(\rm no\, ISW)}_\ell$.}\label{fig:ISWEffect}
\end{figure*}

In Fig.~\ref{fig:DopplerEffect} we show the plots of the fractional change $\Delta{\hat C}_\ell/ C^{(\rm no\, Doppler)}_\ell$ owing to the Doppler correction \eqref{Doppler} in the total magnification angular power spectrum $C_\ell$, as a function of $\ell$, where $\Delta{\hat C}_\ell \,{=}\, C_\ell \,{-}\, C^{(\rm no\, Doppler)}_\ell$. We considered both constant $\alpha_H$ (left panels) and dynamic $\alpha_H$ (right panels) scenarios, \eqref{case1} and \eqref{case2}, respectively, for the same parameters as in Fig.~\ref{fig:totalCls}. Similarly, the plots are given at $z_S \,{=}\, 0.5$ (top), $z_S \,{=}\, 1$ (middle), and $z_S \,{=}\, 3$ (bottom). These fractions measure the Doppler signal in the total magnification angular power spectrum (i.e. the Doppler magnification signal) with respect to the Horndeski parameter, at the given source redshifts. For constant $\alpha_H$ (left panels) we see that at each $z_S$, as the value of the Horndeski parameter increases, the amplitude of the Doppler magnification signal also increases: at $z_S \,{>}\, 0.5$ the growth in amplitude appears to occur on all scales; whereas, at $z_S \,{=}\, 0.5$ the growth occurs on scales $\ell \,{\lesssim}\, 20$, while on scales $\ell \,{>}\, 20$ the behaviour of the Doppler magnification signal reverses, where an increasing $\alpha_H \,{\geq}\, 0.15$ will lead to a decreasing amplitude of the Doppler magnification signal. This implies that stronger UDE ($\alpha_H \,{\geq}\, 0.15$), and hence beyond-Horndeski gravity, will enhance the signal of Doppler effect in the magnification of cosmic objects on all scales at earlier epochs ($z_S \,{>}\, 0.5$), and only on the largest scales ($\ell \,{\lesssim}\, 20$) at late-time epochs ($z_S \,{\leq}\, 0.5$). On the other hand, for dynamic $\alpha_H$ (right panels) we observe a similar behaviour in the Doppler magnification signal, except that here the signal is continuously suppressed for larger $e_H$, at all $z_S$, i.e. the converse of the constant $\alpha_H$ scenario. Thus, for dynamic $\alpha_H$, stronger UDE (larger $e_H$) will suppress Doppler magnification signal (given our choice of parameterisation).

\begin{figure*}[!h]\centering
\includegraphics[scale=0.4]{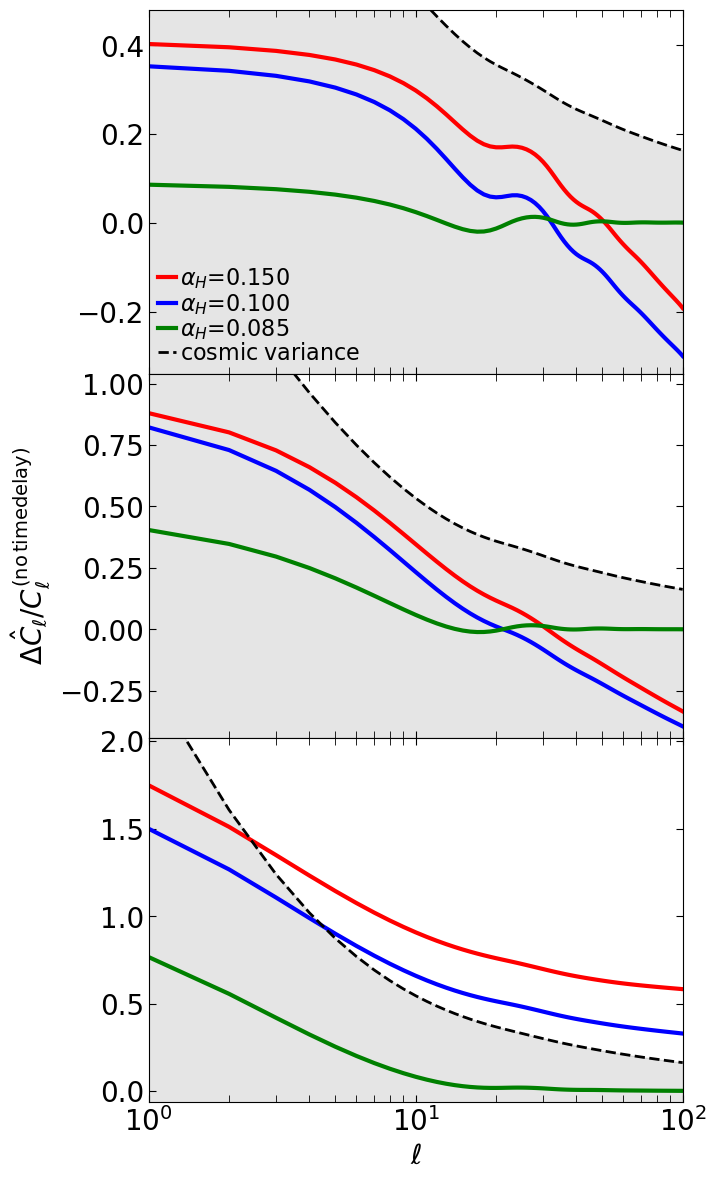} \includegraphics[scale=0.4]{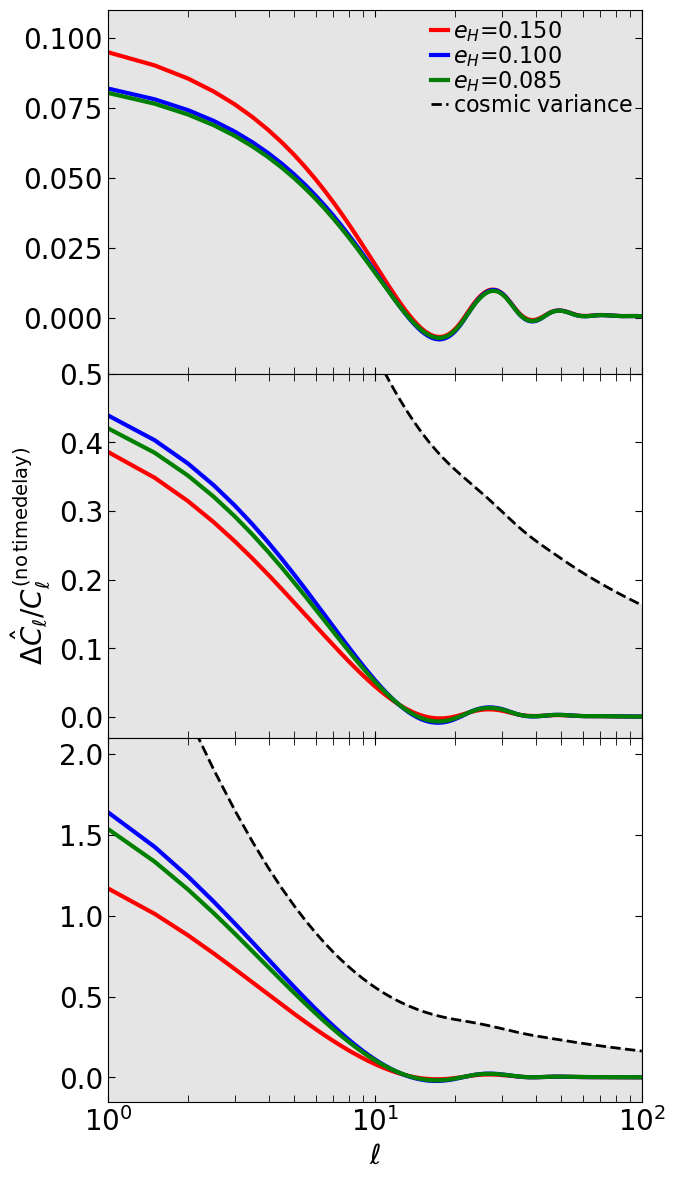} 
\caption{The plots of time-delay signal in the total magnification angular power spectrum $C_\ell$ for the same UDE parameters and scenarios as in Fig.~\ref{fig:totalCls}. \emph{Left:} Plots for constant $\alpha_i$ ($i \neq M$) scenario \eqref{case1}. \emph{Right:} Plots for dynamic $\alpha_i$ scenario \eqref{case2}. Similarly, the plots in the \emph{Left} and \emph{Right} panels at $z_S \,{=}\, 0.5$ ({\it top}), $z_S \,{=}\, 1.0$ ({\it middle}), and $z_S \,{=}\, 3.0$ ({\it bottom}), where $\Delta{\hat C}_\ell \,{=}\, C_\ell \,{-}\, C^{(\rm no\, timedelay)}_\ell$.}\label{fig:timedelayEffect}
\end{figure*}

\begin{figure*}[!h]\centering
\includegraphics[scale=0.4]{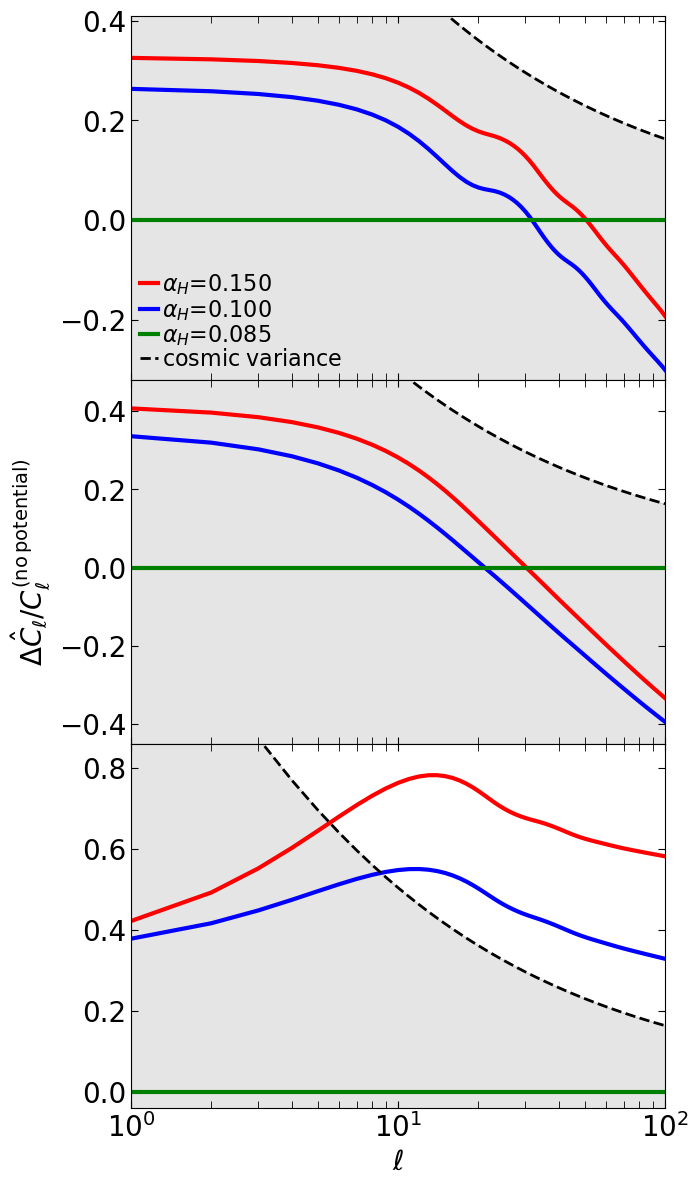} \includegraphics[scale=0.4]{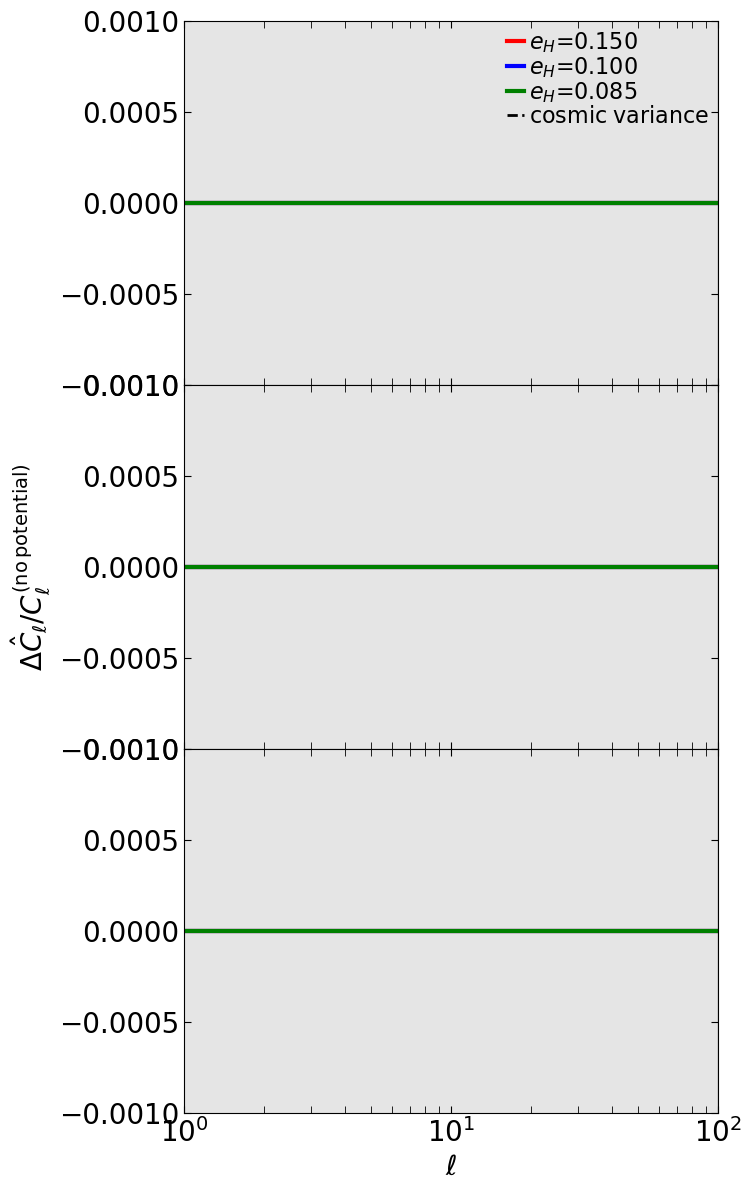} 
\caption{The plots of gravitational (potential) signal in the total magnification angular power spectrum $C_\ell$ for the same UDE parameters and scenarios as in Fig.~\ref{fig:totalCls}. \emph{Left:} Plots for constant $\alpha_i$ scenario \eqref{case1}. \emph{Right:} Plots for dynamic $\alpha_i$ scenario \eqref{case2}. Similarly, the plots in the \emph{Left} and \emph{Right} panels at the source redshifts $z_S \,{=}\, 0.5$ ({\it top}), $z_S \,{=}\, 1.0$ ({\it middle}), and $z_S \,{=}\, 3.0$ ({\it bottom}), where we have $\Delta{\hat C}_\ell \,{=}\, C_\ell \,{-}\, C^{(\rm no\, potential)}_\ell$.}\label{fig:potentialEffect}
\end{figure*}

We also show the extent of the cosmic variance (shaded regions) for the Doppler magnification signal, given by
\beq\label{sigma_eff2}
\sigma^X_\ell = \dfrac{\sqrt{2} }{C^{\rm (no\ X)}_\ell} \sigma_\ell,
\eeq
which follows by similar calculations as in \eqref{sigma_eff}; with $X$ denoting the individual relativistic signals ($X$ = Doppler, ISW, time-delay, potential), and $\sigma_\ell$ is as given by \eqref{sigma}. (Note that $\sigma^X_\ell$ is applied in Figs.~\ref{fig:DopplerEffect}--\ref{fig:potentialEffect}, with $\alpha_H \,{=}\, e_H \,{=}\,0.085$, as in Fig.~\ref{fig:tot2stdClsFracs}.) 

In Fig.~\ref{fig:DopplerEffect}, by going from bottom to top panel (decreasing $z_S$), we notice the behaviour of the Doppler magnification signal with respect to cosmic variance: a gradual growth in amplitude as source redshift decreases, for both constant $\alpha_H$ (left panels) and dynamic $\alpha_H$ (right panels). We see that the Doppler magnification signal gradually surpasses the cosmic variance, at $z_S \,{\lesssim}\, 0.5$ (on different scales $\ell$ for constant $\alpha_H$ and dynamic $\alpha_H$, accordingly). Similarly, this implies that the Doppler magnification signal can be (directly) detected with respect to cosmic variance, i.e. the quantitative analysis of the Doppler signal in the total magnification angular power spectrum will not require the use of multi-tracer methods, at the given $z_S$. Nevertheless, the potential of detecting the Doppler magnificaion signal for dynamic $\alpha_H$ appears to be lower than for constant $\alpha_H$, at all $z_S$ (particularly at $z_S \,{\lesssim}\, 0.5$). Moreover, we see that the separation in the lines of the Doppler magnification signal, between successive values of the Horndeski parameter, increases as source redshift decreases, for both constant $\alpha_H$ and dynamic $\alpha_H$: albeit not obvious for the separation between the green line ($\alpha_H \,{=}\, e_H \,{=}\, 0.085$) and the blue line ($\alpha_H \,{=}\, e_H \,{=}\, 0.1$) at $z_S \,{=}\, 0.5$, which is overshadowed by the separation between the blue line and the red line ($\alpha_H \,{=}\, 0.15$). (For example, for constant $\alpha_H$ the separation between the blue line and the green line is ${\sim}\, 0.01$ at $z_S \,{=}\, 3$, ${\sim}\, 0.1$ at $z_S \,{=}\, 1$, and ${\sim}\, 4$ at $z_S \,{=}\, 0.5$.) Although the values of the separation seem marginal for $\alpha_H \,{=}\, e_H \,{<}\, 0.15$, they nevertheless show that the Doppler magnification signal is relatively more sensitive to changes in UDE at low source redshift ($z_S \,{\lesssim}\, 0.5$). This implies that the Doppler signal in the total magnification angular power spectrum holds the potential of being a probe of the imprint of UDE, at the given source redshift (particularly, for constant $\alpha_H$). 

Fig.~\ref{fig:DopplerEffect} also shows that, for both constant $\alpha_H$ and dynamic $\alpha_H$, the amplitude of the Doppler magnification signal decreases as source redshift increases (top to bottom panels). Therefore, magnification of sources owing to peculiar velocities diminishes at larger cosmic distances. It should be pointed out that the apparent oscillations (on scales $\ell \,{\gtrsim}\, 20$) in the Doppler magnification signal (and other signals), are likely to be from the Bessel spherical function, which tends to dominate the angular power spectrum amplitude, at $z_S \,{\lesssim}\, 1$. Essentially, we see that the behaviour and amplitude of the total relativistic signal in the total magnification angular power spectrum at $z_S \,{\lesssim}\, 0.5$ (Fig.~\ref{fig:tot2stdClsFracs}, top panels) follows that of the Doppler signal, indicating that it is prescribed by the Doppler signal at the given source redshift. Similarly, the results in Fig.~\ref{fig:DopplerEffect}, particularly for dynamic $\alpha_i$, are consistent with those for quintessence by \cite{Mohamed:2025ijc}.

In Fig.~\ref{fig:ISWEffect} we show, for both constant $\alpha_H$ (left panels) and dynamic $\alpha_H$ (right panels), the plots of the fractional change $\Delta{\hat C}_\ell/ C^{(\rm no\, ISW)}_\ell$ owing to the ISW correction \eqref{ISW} in the total magnification angular power spectrum, as a function of $\ell$, where $\Delta{\hat C}_\ell \,{=}\, C_\ell \,{-}\, C^{(\rm no\, ISW)}_\ell$. The plots are given for the same parameters as in Fig.~\ref{fig:totalCls}, at $z_S \,{=}\, 0.5$ (top), $z_S \,{=}\, 1$ (middle), and $z_S \,{=}\, 3$ (bottom). Similarly, we show the extent of cosmic variance (shaded regions), as given by \eqref{sigma_eff2}. These fractions measure the ISW signal in the total magnification angular power spectrum for the given UDE parameter values, at the given source redshifts. For constant $\alpha_H$ (left panels), we see a gradual growth in amplitude in the ISW magnification signal as source redshift increases: at $z_S \,{<}\, 3$ the ISW magnification signal is subdued by cosmic variance on all scales; whereas, at $z_S \,{=}\, 3$, the ISW magnification signal grows and surpasses cosmic variance for $\alpha_H \,{\geq}\, 0.1$, on all scales. (The growth in amplitude with increase in $z_S$ can be understandable since the ISW effect is an integral effect, which will increase with redshift or distance.) Thus, the results imply that the cosmic-variance effect can be neglected in the quantitative analysis of the ISW signal in the total magnification angular power spectrum, for relatively large amplitudes of the Horndeski parameter ($\alpha_H \,{\gtrsim}\, 0.1$), at high source redshifts ($z_S \,{\geq}\, 3$); whereas, at lower source redshifts ($z_S \,{<}\, 3$), multi-tracer methods will need to be incorporated. On the other hand, for dynamic $\alpha_i$ (right panels), we see a gradual growth in amplitude of the ISW magnification signal from $z_S \,{=}\, 0.5$ to $z_S \,{=}\, 1$, and then a decrease at $z_S \,{=}\, 3$. Moreover, similar to the results for quintessence by \cite{Mohamed:2025ijc}, the ISW magnification signal for dynamic $\alpha_i$ is completely surpassed by cosmic variance at all source redshifts. Thus, it will be difficult to isolate the ISW magnification signal without advanced analytical methods, such as multi-tracer analysis.

\begin{figure*}[!h]\centering
\includegraphics[scale=0.4]{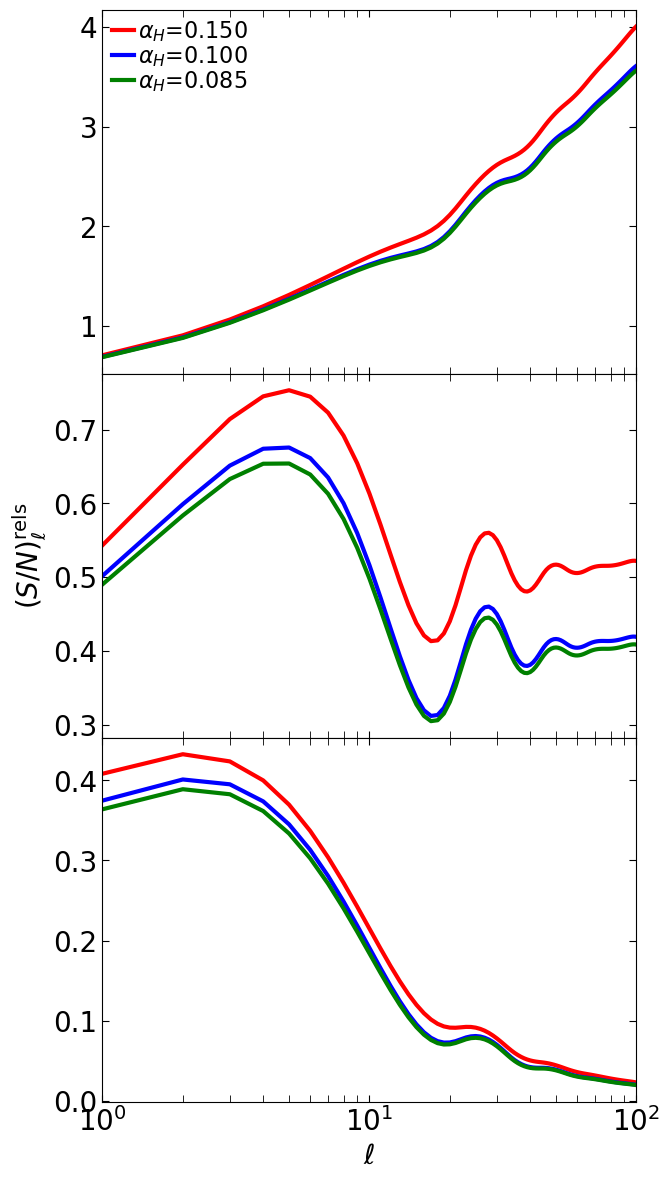} \includegraphics[scale=0.4]{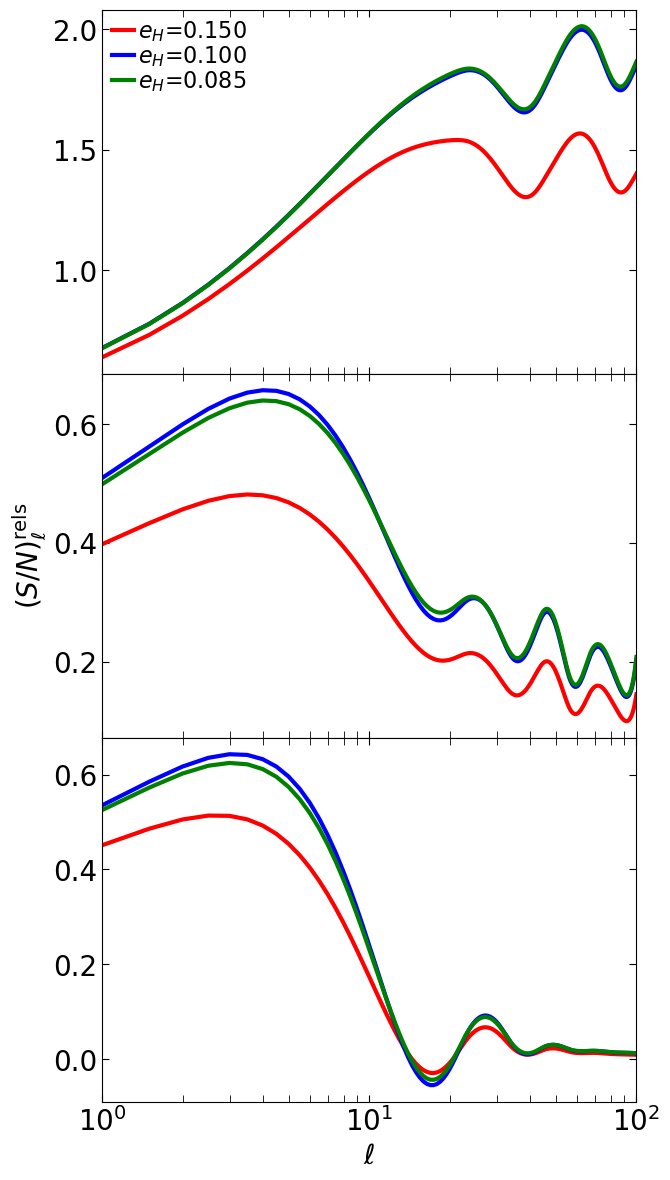}   
\caption{The plots of the total relativistic signal-to-noise ratio in the total magnification angular power spectrum $C_\ell$ for the same UDE parameters and scenarios as in Fig.~\ref{fig:tot2stdClsFracs}. \emph{Left:} Plots for constant $\alpha_i$ ($i \neq M$) scenario \eqref{case1}. \emph{Right:} Plots for dynamic $\alpha_i$ scenario \eqref{case2}. The plots in the \emph{Left} and \emph{Right} panels at $z_S \,{=}\, 0.5$ ({\it top}), $z_S \,{=}\, 1.0$ ({\it middle}), and $z_S \,{=}\, 3.0$ ({\it bottom}). }\label{fig:rels-SN}
\end{figure*}

Furthermore, for both constant $\alpha_H$ and dynamic $\alpha_H$, we see that the separation in the ISW signal in the total magnification angular power spectrum appear to be largest at $z_S \,{=}\, 3$, indicating that the ISW magnification signal is relatively more sensitive to changes in UDE at high source redshift ($z_S \,{\geq}\, 3$). This implies that the ISW signal in the total magnification angular power spectrum holds the potential of being a probe of the imprint of UDE, at larger cosmic distances (higher redshifts). 

Similarly in Fig.~\ref{fig:timedelayEffect} we show, for both constant $\alpha_H$ (left panels) and dynamic $\alpha_H$ (right panels), the plots of the fractional change $\Delta{\hat C}_\ell/ C^{(\rm no\, timedelay)}_\ell$ owing to the time-delay correction \eqref{timedelay} in the total magnification angular power spectrum $C_\ell$, where we have $\Delta{\hat C}_\ell \,{=}\, C_\ell \,{-}\, C^{(\rm no\, timedelay)}_\ell$. The plots are given for the same UDE parameters, at $z_S \,{=}\, 0.5$ (top), $z_S \,{=}\, 1$ (middle), and $z_S \,{=}\, 3$ (bottom). We also show the extent of cosmic variance (shaded regions). These fractions measure the time-delay signal in the total magnification angular power spectrum for the chosen Horndeski parameter values, at the given source redshifts. For constant $\alpha_H$ (left panels) we see that at $z_S \,{<}\, 3$, the behaviour of the time-delay magnification signal is similar to that of the ISW magnification signal (Fig.~\ref{fig:ISWEffect}, left, middle and top), except that the amplitude of the time-delay magnification signal is larger; whereas at $z_S \,{=}\, 3$, the behaviour of the time-delay magnification signal appears to resemble the inverse of the ISW magnification signal somewhat (for $\alpha_H \,{\geq}\, 0.1$). The time-delay magnification signal results also indicate that at $z_S \,{\geq}\, 3$ cosmic-variance effect will be inconsequential in the relevant quantitative analysis. Similar to the ISW magnification signal, the observed growth in amplitude with increase in $z_S$ in the time-delay magnification signal is related to the fact that time delay is an integral effect, which will increase with redshift. Essentially, the results suggest that in analysing the time-delay magnification signal (for constant $\alpha_H$) at high source redshifts ($z_S \,{\geq}\, 3$) advanced analytical methods like those of multi-tracer analysis can suitably be disregarded. Conversely, for dynamic $\alpha_H$, advanced analytical methods such as multi-tracer analysis must be taken into account to beat down cosmic variance, at all source redshifts, in order to stand the chance of isolating the time-delay magnification signal. (This is consistent with results for quintessence by \cite{Mohamed:2025ijc}.)

Moreover, for both constant $\alpha_H$ and dynamic $\alpha_H$, we see the separation in the time-delay signal in the total magnification angular power spectrum increasing with increase in source redshift, with the largest separation occurring at $z_S \,{=}\, 3$. This suggests that the time-delay magnification signal is relatively more sensitive to changes in UDE at high source redshift. Thus, the time-delay signal in the total magnification angular power spectrum holds the potential to detect the UDE imprint, at $z_S \,{\geq}\, 3$. 

\begin{figure*}[!h]\centering
\includegraphics[scale=0.4]{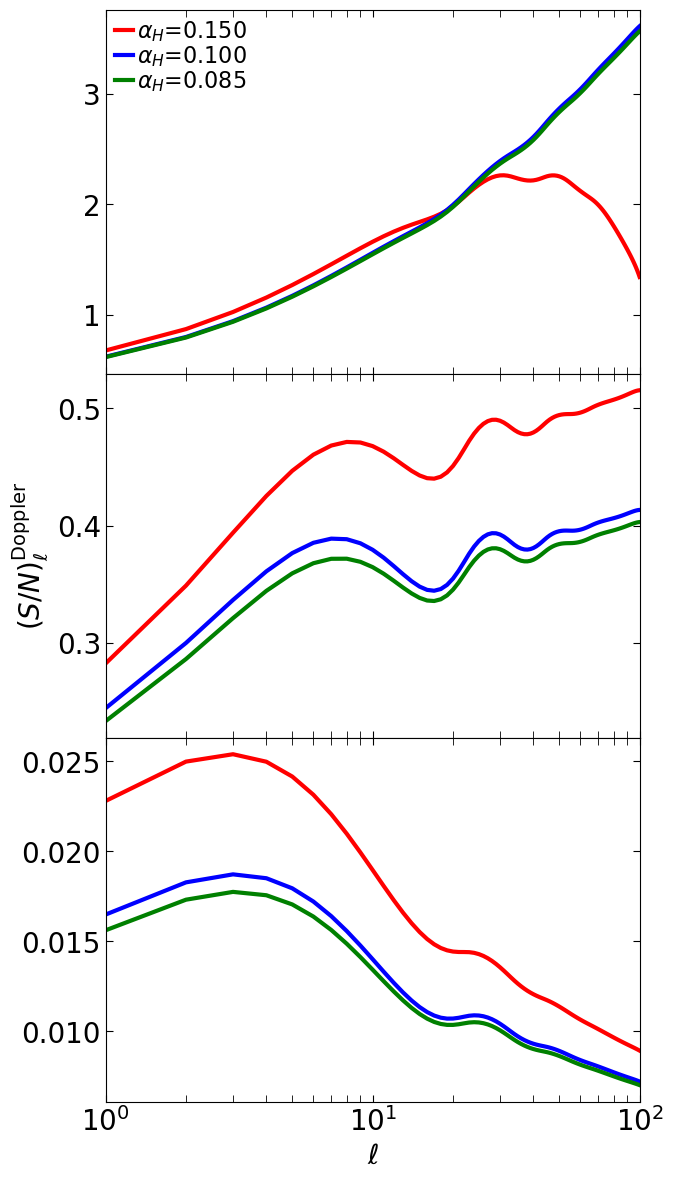} \includegraphics[scale=0.4]{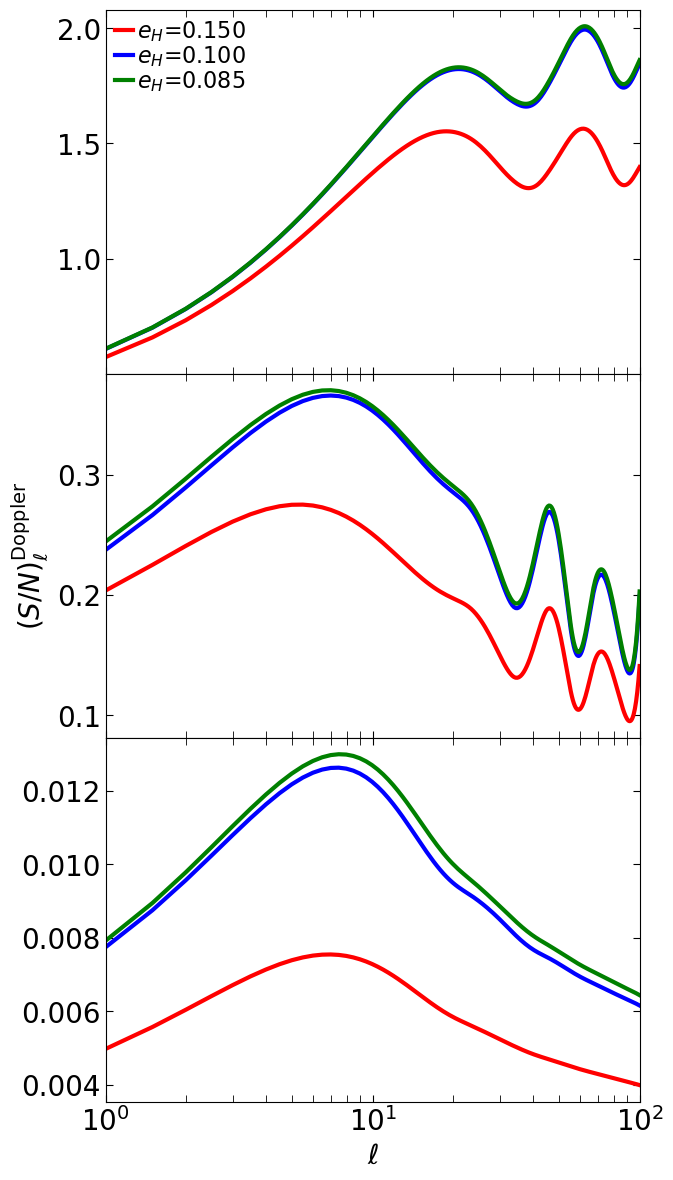} 
\caption{The plots of the Doppler signal-to-noise ratio in the total magnification angular power spectrum $C_\ell$ for the same UDE parameters and scenarios as in Fig.~\ref{fig:DopplerEffect}. \emph{Left:} Plots for constant $\alpha_i$ ($i \neq M$) scenario \eqref{case1}. \emph{Right:} Plots for dynamic $\alpha_i$ scenario \eqref{case2}. Similarly, the plots in the \emph{Left} and \emph{Right} panels at $z_S \,{=}\, 0.5$ ({\it top}), $z_S \,{=}\, 1.0$ ({\it middle}), and $z_S \,{=}\, 3.0$ ({\it bottom}).}\label{fig:Doppler-SN}
\end{figure*}

In Fig.~\ref{fig:potentialEffect} we show, for both constant $\alpha_H$ (left panels) and dynamic $\alpha_H$ (right panels), the plots of the fractional change $\Delta{\hat C}_\ell/ C^{(\rm no\, potential)}_\ell$ owing to the gravitational potential correction \eqref{potentials} in the total magnification angular power spectrum $C_\ell$, where $\Delta{\hat C}_\ell \,{=}\, C_\ell - C^{(\rm no\, potential)}_\ell$. The plots are given for the same UDE parameters as in Fig.~\ref{fig:totalCls}, at $z_S \,{=}\, 0.5$ (top), $z_S \,{=}\, 1$ (middle), and $z_S \,{=}\, 3$ (bottom). We also show the extent of cosmic variance (shaded regions). These fractions measure the gravitational-potential signal in the total magnification angular power spectrum for the given Horndeski parameter values, at the given source redshifts. For constant $\alpha_H$ (left panels), we see that the behaviour of the gravitational (potential) magnification signal is similar to that of the ISW magnification signal, at the given source redshifts, except that the amplitude of the gravitational magnification signal at the given source redshifts is relatively lower. Thus, similar discussion follow for the gravitational magnification signal (as for the ISW magnification signal). Moreover, the gravitational magnification signal appears to tend to vanish for smaller Horndeski parameter values ($\alpha_H \,{\lesssim}\, 0.085$), as opposed to the corresponding ISW magnification signal for the same Horndeski parameter values (the same for the time-delay magnification signal). This is understandable since the ISW magnification and the time-delay signals, respectively, are sourced by integral effects, whose amplitude will accumulate over cosmological distances; on the other hand, the gravitational magnification signal is sourced by local (non-integral) gravitational potential-well effects that are independent of distance, and hence will grow relatively much slower. 

\begin{figure*}[!h]\centering 
\includegraphics[scale=0.4]{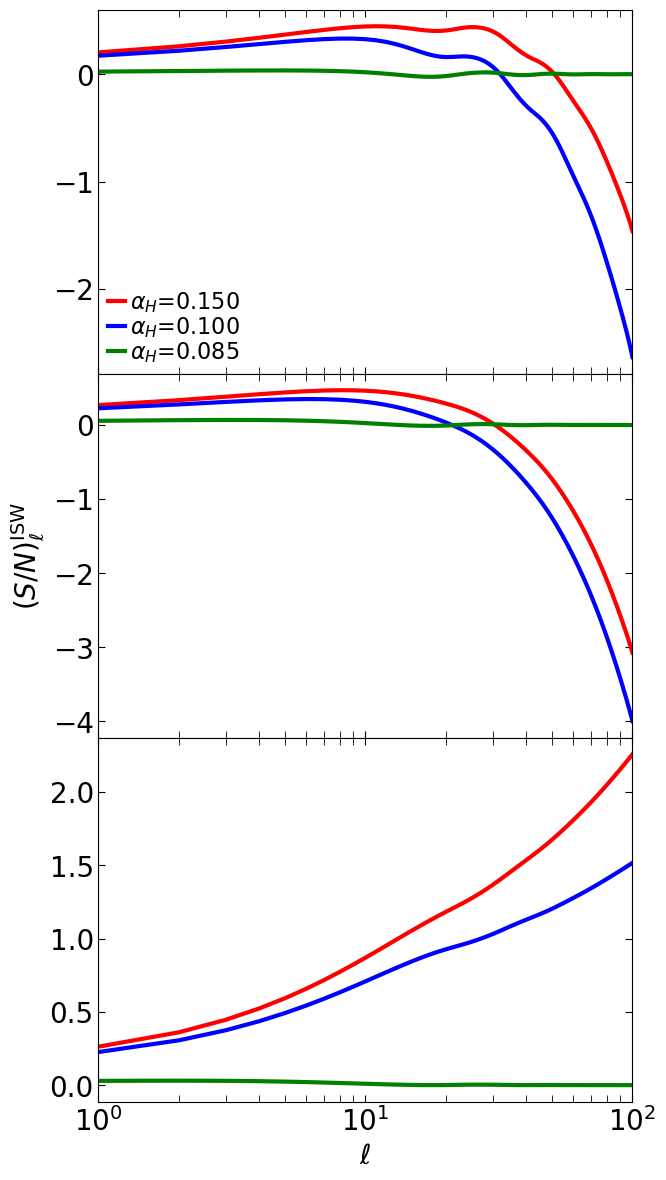} \includegraphics[scale=0.4]{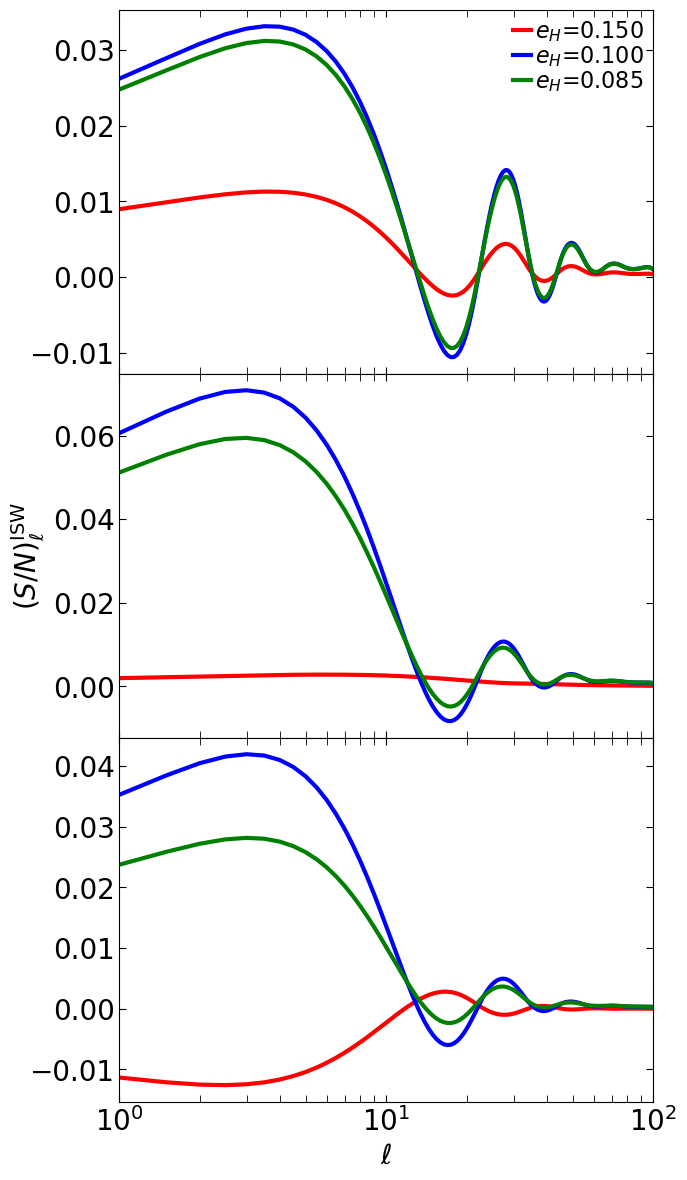} 
\caption{The plots of the ISW signal-to-noise ratio in the total magnification angular power spectrum $C_\ell$ for the same UDE parameters and scenarios as in Fig.~\ref{fig:ISWEffect}. \emph{Left:} Plots for constant $\alpha_i$ ($i \neq M$) scenario \eqref{case1}. \emph{Right:} Plots for dynamic $\alpha_i$ scenario \eqref{case2}. Similarly, the plots in the \emph{Left} and \emph{Right} panels at $z_S \,{=}\, 0.5$ ({\it top}), $z_S \,{=}\, 1.0$ ({\it middle}), and $z_S \,{=}\, 3.0$ ({\it bottom}).}\label{fig:ISW-SN}
\end{figure*}

In general, for constant $\alpha_H$ the results show that the Doppler magnification signal dominates at low redshifts ($z \,{\lesssim}\, 0.5$) while the ISW magnification, the time-delay magnification, and the gravitational magnification signals, respectively, dominate at high redshifts ($z \,{\gtrsim}\, 3$). This is consistent with the results found in e.g. \cite{Duniya:2022xcz, Bonvin:2014owa}. Conversely, for dynamic $\alpha_H$, the Doppler magnification signal also dominates at low redshifts while the ISW magnification, the time-delay magnification, and the gravitational magnification signals, respectively, become surpassed by cosmic variance at all redshifts; with the gravitational magnification signal vanishing identically, at all redshifts. These are consistent with results for quintessence by \cite{Mohamed:2025ijc}. For completeness we give the associated signal-to-noise ratios of all signals in \S\ref{sec:SN-ratio}.

\section{Signal-to-Noise Ratio}\label{sec:SN-ratio}

Here we consider a (theoretical) signal-to-noise ratio ($S/N$) for results discussed in \S\ref{subsec:UDE-imprint} and~\S\ref{subsec:rels-effects}. As previously stated, for the purpose of this work we only consider $S/N$ with respect to cosmic variance. The estimations of $S/N$ will help in shedding light on the possibility of detecting the magnification signals in observational data.

By using $\sigma^{\rm rels}_\ell$ \eqref{sigma_eff}, we obtain the signal-to-noise ratio for the total relativistic signal in the total magnification angular power spectrum (Fig.~\ref{fig:tot2stdClsFracs}), given by
\beq\label{SN-rels}
\left(\dfrac{S}{N}\right)^{\rm rels}_\ell = \dfrac{1}{2} \sqrt{(2\ell + 1)f_{\rm sky}} \left(1 - \dfrac{C^{\rm std}_\ell}{C_\ell}\right),
\eeq
where $C^{\rm std}_\ell$ and $C_\ell$ are as given in \S\ref{sec:Cls_mag}. For a signal to be significant, we need $(S/N)^{\rm rels}_\ell \,{>}\, 1$.

In Fig.~\ref{fig:rels-SN} we show, for both constant $\alpha_H$ (left panels) and dynamic $\alpha_H$ (right panels), the plots of the signal-to-noise ratio $(S/N)^{\rm rels}_\ell$ of the total relativistic signal in the total magnification angular power spectrum, as a function of multipole $\ell$: for the same UDE parameters as in Fig.~\ref{fig:tot2stdClsFracs}, at $z_S \,{=}\, 0.5$ (top), $z_S \,{=}\, 1.0$ (middle), and $z_S \,{=}\, 3.0$ (bottom). For both constant $\alpha_H$ and dynamic $\alpha_H$, we see that at $z_S \,{=}\, 0.5$, we have $(S/N)^{\rm rels}_\ell \,{>}\, 1$ and increases on scales $\ell \,{\gtrsim}\, 3$; whereas, at $z_S \,{\geq}\, 1$, we have $(S/N)^{\rm rels}_\ell \,{\leq}\, 1$. Thus, this suggests that the total relativistic signal in the total magnification angular power spectrum is indeed significant with respect to the cosmic variance, at low source redshifts ($z_S \,{<}\, 1$); whereas at higher source redshifts ($z_S \,{\geq}\, 1$), surveys will require multi-tracer methods to beat down cosmic variance, before the total relativistic magnification signal can become significant. This is consistent with the results in \S\ref{subsec:UDE-imprint} (see Fig.~\ref{fig:tot2stdClsFracs}). 

Similarly, by using $\sigma^X_\ell$ \eqref{sigma_eff2}, we obtain the $S/N$ for the individual relativistic signals in the total magnification angular power spectrum (see Figs.~\ref{fig:DopplerEffect}--\ref{fig:potentialEffect}), given by
\beq\label{SN-relsX}
\left(\dfrac{S}{N}\right)^{\rm X}_\ell = \dfrac{1}{2} \sqrt{(2\ell + 1)f_{\rm sky}} \left(1 - \dfrac{C^{(\rm no\, X)}_\ell}{C_\ell}\right),
\eeq
where $C^{(\rm no\, X)}_\ell$ and $C_\ell$ are as given in \S\ref{sec:Cls_mag}; with $X$ denoting the individual relativistic signals. In essence, $(S/N)^{\rm X}_\ell$ is a generalisation of $(S/N)^{\rm rels}_\ell$ (just as $\sigma^X_\ell$ generalises $\sigma^{\rm rels}_\ell$). 

\begin{figure*}[!h]\centering 
\includegraphics[scale=0.4]{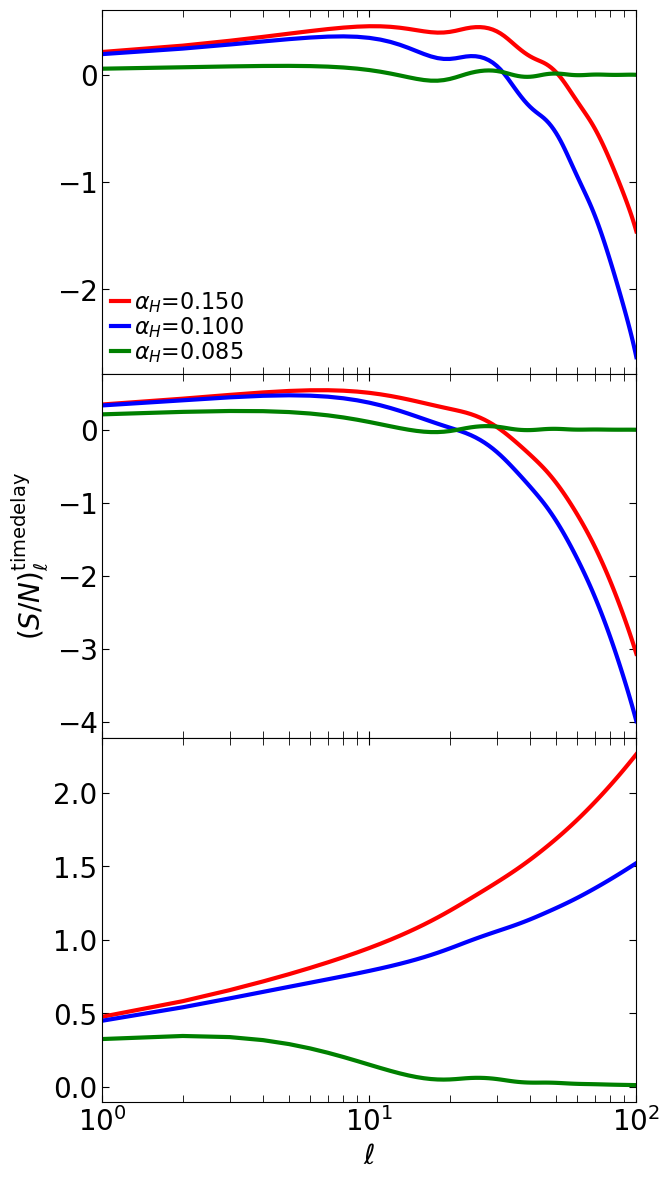} \includegraphics[scale=0.4]{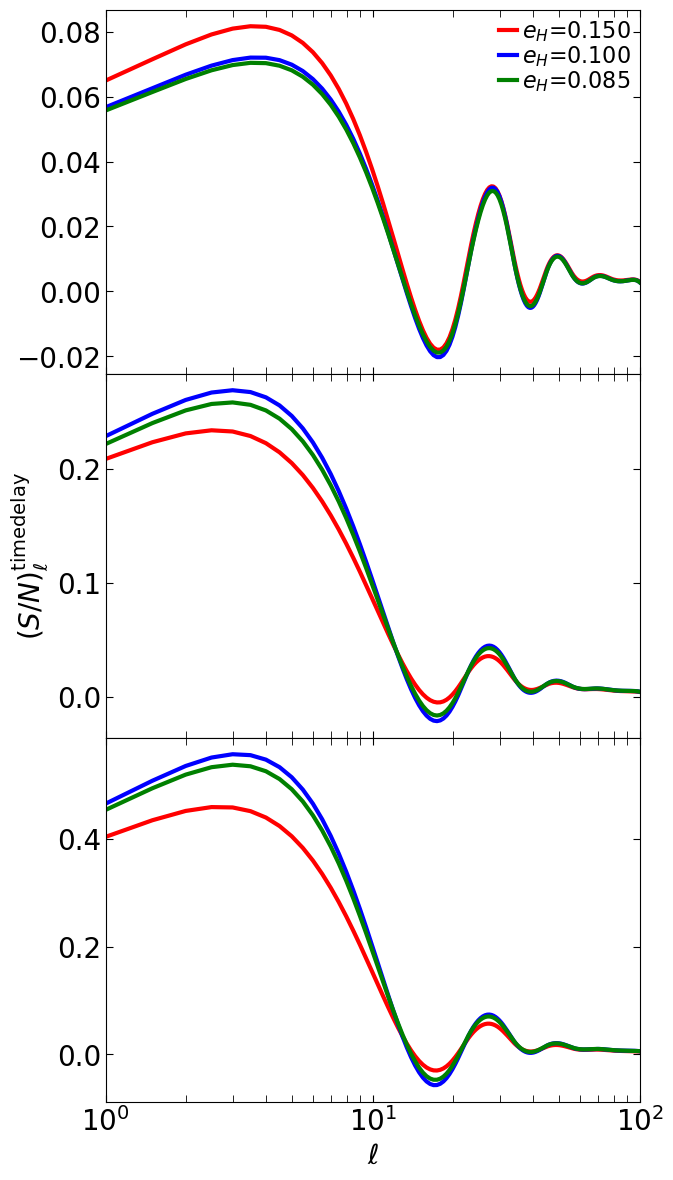} 
\caption{The plots of the time-delay signal-to-noise ratio in the total magnification angular power spectrum $C_\ell$ for the same UDE parameters and scenarios as in Fig.~\ref{fig:timedelayEffect}. \emph{Left:} Plots for constant $\alpha_i$ ($i \neq M$) scenario \eqref{case1}. \emph{Right:} Plots for dynamic $\alpha_i$ scenario \eqref{case2}. Similarly, the plots in the \emph{Left} and \emph{Right} panels at $z_S \,{=}\, 0.5$ ({\it top}), $z_S \,{=}\, 1.0$ ({\it middle}), and $z_S \,{=}\, 3.0$ ({\it bottom}).}\label{fig:timedelay-SN}
\end{figure*}

In Fig.~\ref{fig:Doppler-SN} we show, for both constant $\alpha_H$ (left panels) and dynamic $\alpha_H$ (right panels), the plots of the Doppler signal-to-noise ratio $(S/N)^{\rm Doppler}_\ell$ in the total magnification angular power spectrum $C_\ell$, as a function of multipole $\ell$: for the same UDE parameters as in Fig.~\ref{fig:DopplerEffect}, at $z_S \,{=}\, 0.5$ (top), $z_S \,{=}\, 1.0$ (middle), and $z_S \,{=}\, 3.0$ (bottom). For constant $\alpha$ (left panels), we see that at $z_S \,{=}\, 0.5$, $(S/N)^{\rm Doppler}_\ell \,{>}\, 1$ and increases on scales $\ell \,{\gtrsim}\, 3$ for $\alpha_H \,{<}\, 0.15$---with that of $\alpha_H \,{=}\, 0.15$ decreasing on scales $\ell \,{>}\, 20$; whereas, at $z_S \,{\geq}\, 1$, we have $(S/N)^{\rm Doppler}_\ell \,{\leq}\, 1$. The behaviour of $(S/N)^{\rm Doppler}_\ell$ for $\alpha_H \,{=}\, 0.15$, at $z_S \,{=}\, 0.5$, can be understood from the behaviour of the Doppler magnification signal in \S\ref{subsec:rels-effects} (see Fig.~\ref{fig:DopplerEffect}), at the given $z_S$: the amplitude of the Doppler magnification signal for $\alpha_H \,{=}\, 0.15$ diminishes relative to that of lower values of $\alpha_H$, on scales $\ell \,{\gtrsim}\, 20$. Similar general behaviours are observed for dynamic $\alpha_H$ (right panels), where $(S/N)^{\rm Doppler}_\ell \,{>}\, 1$ at $z_S \,{\lesssim}\, 0.5$, and $(S/N)^{\rm Doppler}_\ell \,{<}\, 1$ at $z_S \,{>}\, 0.5$. Thus, the behaviour of $(S/N)^{\rm Doppler}_\ell$ suggests that the Doppler magnification signal is generally cosmologically significant relative to the cosmic variance, at low source redshifts ($z_S \,{\lesssim}\, 0.5$); whereas at higher source redshifts ($z_S \,{\geq}\, 1$), multi-tracer techniques will need to be taken into account, to beat down cosmic variance to reveal any possible relative significance. This is consistent with the results in \S\ref{subsec:rels-effects}. The behaviour of $(S/N)^{\rm Doppler}_\ell$ for $\alpha_H \,{=}\, 0.15$ at $z_S \,{=}\, 0.5$, may be an indication that the occurrence of Doppler effect in cosmic magnification in beyond-Horndeski gravity may rather prefer a relatively weaker gravity regime ($\alpha_H \,{<}\, 0.15$), on scales $\ell \,{>}\, 20$.

\begin{figure*}[!h]\centering 
\includegraphics[scale=0.4]{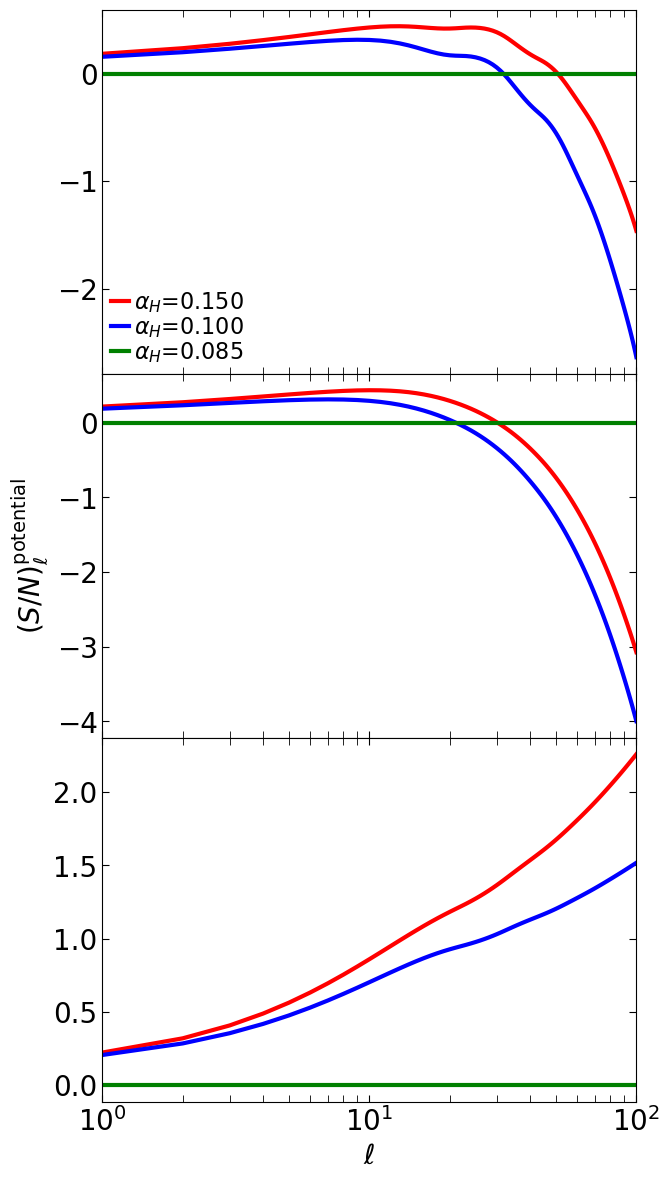} \includegraphics[scale=0.4]{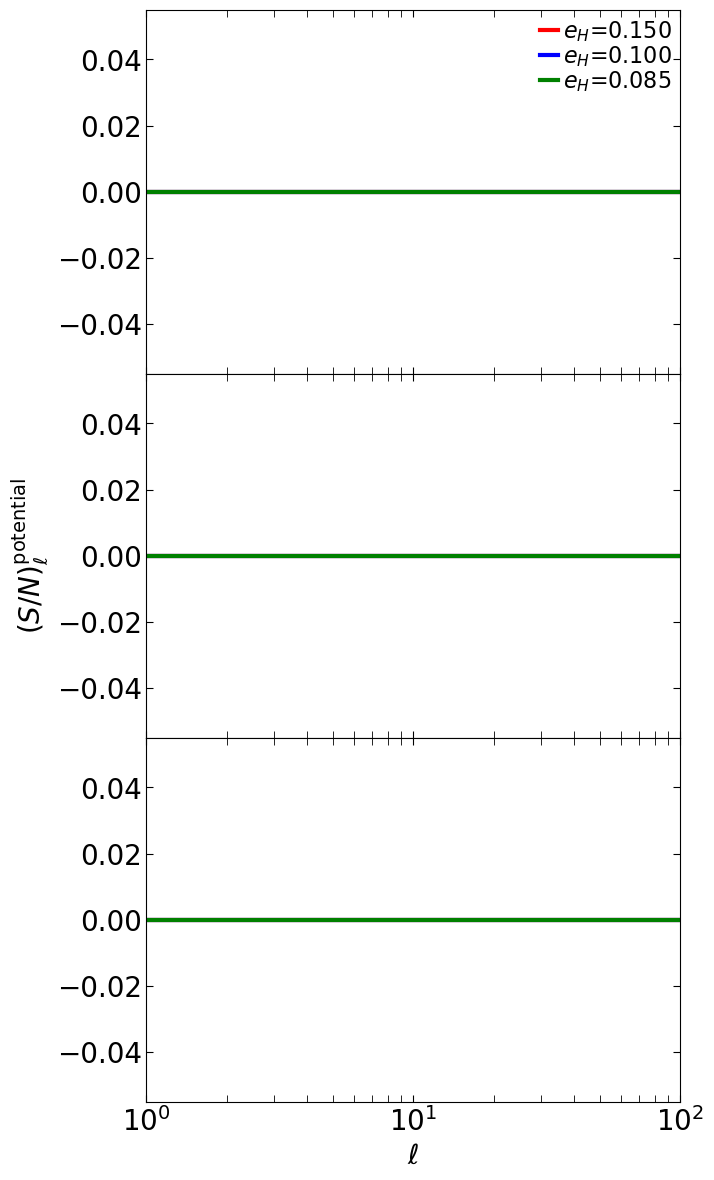} 
\caption{The plots of the gravitational (potential) signal-to-noise ratio in the total magnification angular power spectrum $C_\ell$ for the same UDE parameters and scenarios as in Fig.~\ref{fig:potentialEffect}. \emph{Left:} Plots for constant $\alpha_i$ ($i \neq M$) scenario \eqref{case1}. \emph{Right:} Plots for dynamic $\alpha_i$ scenario \eqref{case2}. Similarly, the plots in the \emph{Left} and \emph{Right} panels at $z_S \,{=}\, 0.5$ ({\it top}), $z_S \,{=}\, 1.0$ ({\it middle}), and $z_S \,{=}\, 3.0$ ({\it bottom}).}\label{fig:potential-SN}
\end{figure*}

In Fig.~\ref{fig:ISW-SN} we show, for both constant $\alpha_H$ (left panels) and dynamic $\alpha_H$ (right panels), the plots of the ISW signal-to-noise ratio $(S/N)^{\rm ISW}_\ell$ in the total magnification angular power spectrum $C_\ell$, as a function of multipole $\ell$: for the same UDE parameters as in Fig.~\ref{fig:ISWEffect}, at $z_S \,{=}\, 0.5$ (top), $z_S \,{=}\, 1.0$ (middle), and $z_S \,{=}\, 3.0$ (bottom). For constant $\alpha_H$ (left panels), we see that at $z_S \,{<}\, 3$ we have $(S/N)^{\rm ISW}_\ell \,{\leq}\, 1$, and at $z_S \,{=}\, 3$, we have $(S/N)^{\rm ISW}_\ell \,{>}\, 1$ and increases for $\alpha_H \,{\geq}\, 0.1$ on scales $\ell \,{\gtrsim}\, 10$. Similarly, this suggests that for constant $\alpha_H$ the ISW signal in the total magnification angular power spectrum will be cosmologically significant relative to the cosmic variance, at high redshifts ($z \,{\geq}\, 3$); whereas at lower redshifts, the analysis will require multi-tracer methods to beat down cosmic variance. Conversely, for dynamic $\alpha_H$ (right panels) we have $(S/N)^{\rm ISW}_\ell \,{<}\, 1$ for all $z_S$: implying that the ISW magnification signal will not be directly observable owing to cosmic variance, at all source redshifts, for dynamic $\alpha_H$. This is consistent with the results in Fig.~\ref{fig:ISWEffect}. 

In Fig.~\ref{fig:timedelay-SN} we show, for both constant $\alpha_H$ (left panels) and dynamic $\alpha_H$ (right panels), plots of the time-delay signal-to-noise ratio $(S/N)^{\rm timedelay}_\ell$ in the total magnification angular power spectrum $C_\ell$, as a function of multipole $\ell$: for the same UDE parameters as in Fig.~\ref{fig:timedelayEffect}, at $z_S \,{=}\, 0.5$ (top), $z_S \,{=}\, 1.0$ (middle), and $z_S \,{=}\, 3.0$ (bottom). For constant $\alpha_H$ (left panels), we see that we have $(S/N)^{\rm timedelay}_\ell \,{\leq}\, 1$ at $z_S \,{<}\, 3$; whereas, at $z_S \,{=}\, 3$ we have $(S/N)^{\rm timedelay}_\ell \,{>}\, 1$ and increases on scales $\ell \,{\gtrsim}\, 10$ for $\alpha_H \,{\gtrsim}\, 0.1$. (This is similar to $(S/N)^{\rm ISW}_\ell$.) This suggests that the time-delay magnification signal will be significant relative to the cosmic variance in the observational data at high redshifts ($z \,{\geq}\, 3$), for strong gravity regimes with coonstant $\alpha_H$. At low redshifts ($z \,{<}\, 3$), however, the time-delay signal will only become cosmologically significant by incorporating multi-tracer methods. Conversely, for dynamic $\alpha_H$ (right panels) we have $(S/N)^{\rm timedelay}_\ell \,{<}\, 1$ for all $z_S$, and hence implying that the time-delay magnification signal will require advanced methods like multi-tracer analysis, at all source redshifts, in order to be observable. This is consistent with the results in Fig.~\ref{fig:timedelayEffect}. 

Similarly in Fig.~\ref{fig:potential-SN} we show, for both constant $\alpha_H$ (left panels) and dynamic $\alpha_H$ (right panels), the plots of the (gravitational) potential signal-to-noise ratio $(S/N)^{\rm potential}_\ell$ in the total magnification angular power spectrum $C_\ell$, as a function of multipole $\ell$: for the same UDE parameters as in Fig.~\ref{fig:timedelayEffect}, at $z_S \,{=}\, 0.5$ (top), $z_S \,{=}\, 1.0$ (middle), and $z_S \,{=}\, 3.0$ (bottom). For constant $\alpha_H$ (left panels), we see that we have almost identical results as for $(S/N)^{\rm timedelay}_\ell$, except that for $\alpha_H \,{=}\, 0.085$ the potential magnification signal-to-noise ratio vanishes, $(S/N)^{\rm potential}_\ell \,{=}\, 0$ at all $z_S$. Moreover, this is consistent with the results in Fig.~\ref{fig:potentialEffect} (left panels). Thus, similar deductions as for $(S/N)^{\rm timedelay}_\ell$ follow. It appears that at $z_S \,{\geq}\, 3$, the ISW, the time-delay, and the gravitational magnification signals, respectively, will prefer stronger gravity regimes ($\alpha_H \,{\gtrsim}\, 0.1$) with respect to cosmic variance; otherwise, the signals become diminished. On the other hand, for dynamic $\alpha_H$ (right panels), we have that the gravitational magnification signal-to-noise ratio is identically zero, at all source redshifts: consistent with results in Fig.~\ref{fig:potentialEffect} (right panels).

\section{Conclusion}\label{sec:Concl}

We presented a qualitative analysis of cosmic magnification in the beyond-Horndeski gravity, for both constant phenomenology and dynamic phenomenology, via the unified dark energy model. We used the angular power spectrum for our analysis. We set the effective equation of state parameter of the given unified dark energy to a value which allows the recovery of the background cosmology of general relativity in the matter-dominated era, e.g. before cosmic acceleration started. Moreover, the chosen values of the relevant parameters were such that the same values of $H_0$ and $\Omega_{m0}$ were obtained. This ensured that any deviations in the magnification angular power spectrum that are solely owing to the perturbations in the unified dark energy (hence, beyond-Horndeski gravity) were isolated on large scales.

We discussed the imprint of unified dark energy and the total (relativistic) magnification angular power spectrum. We investigated the total signal of relativistic effects, and the signal of the individual relativistic effects, in the total magnification angular power spectrum. We also estimated the (theoretical) signal-to-noise ratio of various magnification signals. For the purpose of this work, we only consider cosmic variance as the source of error. Furthermore, we consider a best-case scenario for the cosmic variance by adopting an SKA2-like value of the sky-coverage fraction $f_{\rm sky}=0.75$, which is sufficient to achieve the purpose of the paper. 

For both constant phenomenology and dynamic phenomenology, our results showed that in the beyond-Horndeski theory, strong gravity regimes will induce strong cosmic magnification events, and vice versa. Moreover, we found that advanced methods such as multi-tracer analysis (see e.g.~\cite{Abramo:2013awa, Alonso:2015sfa, Fonseca:2015laa, Witzemann:2018cdx, Paul:2022xfx, Karagiannis:2023lsj, Zhao:2023ebp}) will be needed to beat down cosmic variance, at all redshifts, in order for the total cosmic magnification signal to be cosmologically significant; and in principle, increase the possibility of its detection (taking proper quantitative methods into account). For the total relativistic (non-lensing) magnification signal and the Doppler magnification signal, our analysis suggests that the angular power spectrum holds the potential of isolating these signals, at low redshifts ($z \,{\lesssim}\, 0.5$). However, at higher redshifts ($z \,{>}\, 0.5$), multi-tracer analysis will be required to beat down cosmic variance on scales $\ell \,{>}\, 3$, before they can be detectable. This suggests that analysis of magnification surveys (e.g. cosmological surveys that measure angular size of sources) for either the Doppler or the total relativistic magnification signal, at redshifts $z \,{\lesssim}\, 0.5$, will not require the advanced methods of multi-tracer analysis in beyond-Horndeski gravity. (Prospects of measuring the relativistic angular distance or convergence, and hence cosmic magnification, with the SKA has been explored by e.g. \cite{Bonvin:2008ni, Zhang:2005pu, Zhang:2005eb}. See also~\cite{Weltman:2018zrl}, for prospects of radio weak-lensing and Doppler magnification measurements with the SKA.)

Furthermore, unlike in the Doppler magnification signal where the amplitude increased with decreasing redshift, the amplitude of the ISW magnification, time-delay magnification, and gravitational (potential) magnification signals, respectively, increased with increasing redshift, and become cosmologically significant (with respect to the cosmic variance, at the given redshifts). The individual relativistic magnification signals are sensitive to small changes in unified dark energy (at different redshifts): this will be crucial in identifying the imprint of beyond-Horndeski gravity in the magnification angular power spectrum.

The Doppler magnification signal dominated in the beyond-Horndeski gravity, at low redshifts ($z \,{\lesssim}\, 0.5$) for both constant and dynamic phenomenologies, respectively; whereas, the ISW, the time-delay, and gravitational magnification signals, respectively, dominated at high redshifts ($z \,{\gtrsim}\, 3$) for constant phenomenology. For dynamic phenomenology, only the ISW and the time-delay magnification signals showed the potential for growth at $z \,{\geq}\, 3$, with the gravitational (potential) magnification signal vanishing identically at all redshifts. The features exhibited by results for dynamic phenomenology were found to be consistent with results for quintessence \cite{Mohamed:2025ijc}. The signal-to-noise ratios of all these magnification signals---for both constant and dynamic phenomenologies, respectively---were found to be consistent with the behaviour of the respective signals. The results further suggested that for constant phenomenology, the Doppler magnification signal in beyond-Horndeski gravity will prefer a relatively weaker gravity regime ($\alpha_H \,{<}\, 0.15$), and the other signals will prefer stronger gravity regimes ($\alpha_H \,{\gtrsim}\, 0.1$). Conversely, for dynamic phenomenology, a relatively weaker gravity strength ($\alpha_H \,{<}\, 0.15$) appeared to be preferable throughout.

\section*{Acknowledgements}
We thank the referee for useful comments. We also thank the Centre for High Performance Computing, Cape Town, South Africa, for providing the computing facilities with which all the numerical computations in this work were done. 

\section*{Data Availability}
Data sharing is not applicable to this article, as no datasets were generated or analysed in the current study. 

\appendix

\section{The Cosmological Equations}
\label{sec:CEqs}
The equations given in this appendix are drawn from the works by \cite{Gleyzes:2014rba} and \cite{Duniya:2019mpr}.

\subsection{The Perturbations Equations}
The gravitational potential are given, via the metric \eqref{metric}, by 
\beq
\Phi \;\equiv\; \delta{N} + {\cal H}\Pi + \Pi',\quad \Psi \;\equiv\; -\zeta - {\cal H}\Pi,\quad \Pi \;=\; a\psi,
\eeq
where $\psi$ is a metric scalar potential, $\delta{N}$ is the metric temporal perturbation, and $\zeta$ is a metric spatial potential. The evolutions of the UDE momentum density and (energy) density perturbation---which were used in obtaining \eqref{VxEvoln} and \eqref{DxEvoln}---are given by 
\begin{align}
q'_x + 4{\cal H}q_x + \left(\bar{\rho}_x + \bar{p}_x\right)\Phi + \delta{p}_x - \dfrac{2}{3} k^2 \sigma_x =&\; \alpha_M {\cal H}q,\\
\delta{\rho}'_x + 3{\cal H}\left(\delta{\rho}_x + \delta{p}_x\right) -3\left(\bar{\rho}_x + \bar{p}_x\right) \Psi' - k^2q_x  =&\; \alpha_M {\cal H} \delta\rho,
\end{align}
where $q_A \,{=}\, \left(\bar{\rho}_A+\bar{p}_A\right)V_A$ and $\sigma_A$ are the momentum densities and the anisotropic stress potentials, respectively; with $V_x$ and $\sigma_x$ being as given by \eqref{V_x} and \eqref{sigma_x}, respectively. The UDE pressure perturbation is given by
\begin{align} \label{delRho_x}
\delta{\rho}_x \;{\equiv}\;& 2 k^2\left( \alpha_H{\cal R} - \alpha_B a^{-2} M^2{\cal H}\Pi\right)  + \left(\alpha_K - 6\alpha_B\right) {\cal H}^2 {\cal P} \nn
& - 3{\cal H}\left[ (\bar{\rho}_x+\bar{p}_x)\Pi - 2\alpha_B {\cal Q}\right], 
\end{align}
and,
\begin{align}\label{delP_x}
\delta{p}_x \;{\equiv}\;& \left(\dfrac{\bar{\rho}_x+\bar{p}_x}{a^{-2}M^2} + 6\alpha_B {\cal H}^2\right) {\cal P} - 2\alpha_M {\cal H} {\cal Q} + \dfrac{2}{3} k^2\sigma_x \nn
& +\; \left[\bar{p}'_x + \alpha_M {\cal H} a^{-2}M^2 \left(2{\cal H}' + {\cal H}^2\right) \right]\Pi \nn
& +\; 2\alpha_B \left(1 + \dfrac{\alpha'_B}{{\cal H} \alpha_B} + \dfrac{{\cal H}'}{{\cal H}^2} + \dfrac{{\cal P}'}{{\cal H} {\cal P}}\right) {\cal H}^2 {\cal P},
\end{align}
where,
\begin{align}\label{PRQ}
{\cal P} \;{\equiv}\;& \dfrac{M^2}{a^2} \left(\Pi' + {\cal H}\Pi - \Phi\right),\;\; {\cal R} \equiv \dfrac{M^2}{a^2} \left(\Psi + {\cal H}\Pi\right), \\
{\cal Q} \;{\equiv}\;& \dfrac{M^2}{a^2} \left[\Psi' + {\cal H}\Phi + ({\cal H}' - {\cal H}^2)\Pi\right] .
\end{align}
For the sake of completeness, we retain in the rest of this appendix the pressure-related parameters for matter.


\subsection{The Metric Potentials Evolution Equations}
The evolution equations for the metric potentials $\Pi$ and $\Psi$ are given by
\begin{align}\label{dpi-dt}
\Pi' + \Big(1 + \dfrac{\alpha_T - \alpha_M}{\alpha_H}\Big) {\cal H}\Pi =&\; \Big(\dfrac{1+\alpha_H}{\alpha_H}\Big) \Phi - \Big(\dfrac{1+\alpha_T}{\alpha_H}\Big) \Psi \nn
& +\; 8{\pi}G_{\rm eff}a^2 \dfrac{\sigma_m}{\alpha_H},
\end{align}
where $G_{\rm eff}$ is as given by \eqref{Friedmann}, and
\begin{align}\label{dPsidtApp}
\Psi' + (1+\alpha_B) {\cal H}\Phi =& \left[\alpha_B - \gamma_{\cal H} - \dfrac{4{\pi}G_{\rm eff}a^2}{{\cal H}^2} (\bar{\rho}_m + \bar{p}_m)\right] {\cal H}^2\Pi \nn
& +\; \alpha_B {\cal H}\Pi' - 4{\pi}G_{\rm eff}a^2 q_m ,
\end{align}
where $\gamma_{\cal H} \equiv {\cal H}'/{\cal H}^2 - 1$, and we have 
\begin{align}\label{ddpiddt}
\Pi'' + (1 + \gamma_1) {\cal H}\Pi' +&\; \gamma_3 {\cal H}^2 \Pi - \Phi' + \gamma_4\Psi' + \gamma_5 {\cal H}\Phi + \gamma_6 {\cal H}\Psi \nn
= -&\; \dfrac{8{\pi}G_{\rm eff}a^2}{\gamma_0 {\cal H}} \left( 2 k^2 \sigma_m + 3 \delta{p}_m \right) \alpha_B,
\end{align}
with $\gamma_0 \equiv \alpha_K + 6\alpha_B^2$.

From \S\ref{sec:UDE}, the parameters $\gamma_1$ and $\gamma_3$ are given by
\begin{align}\label{g1}
\gamma_1 \equiv &\; 3+\alpha_M + \dfrac{6\alpha_B}{\gamma_0}\Big[\dfrac{\alpha'_B}{{\cal H}} - \dfrac{4{\pi}G_{\rm eff}a^2}{{\cal H}^2}(\bar{\rho}_m + \bar{p}_m)\Big] \nn
& +\; \dfrac{\alpha'_K}{{\cal H}\gamma_0} + (6\alpha^2_B + 2\alpha_K - 6\alpha_B) \dfrac{\gamma_{\cal H}}{\gamma_0} ,\\
\gamma_3 \equiv &\; \gamma_1 + \gamma_2 + \gamma_{\cal H} + 1,
\end{align}
where $\gamma_{\cal H}$ is as given by \eqref{dPsidtApp}, and
\begin{align}
\gamma_2 \equiv &\; 6\left[\dfrac{\alpha'_B}{{\cal H}} + (1+\alpha_B) \gamma_{\cal H} + \dfrac{4{\pi}G_{\rm eff}a^2}{{\cal H}^2}(\bar{\rho}_m + \bar{p}_m) \right] \dfrac{\gamma_{\cal H}}{\gamma_0} \nn
&\; -\dfrac{2k^2}{\gamma_0{\cal H}^2} \Big[1 + \alpha_T + \alpha_B(1+\alpha_B) - (1+\alpha_H)(1+\alpha_M) \nn
&\; \hspace{1.3cm} +   \dfrac{\alpha'_B - \alpha'_H}{{\cal H}} + \dfrac{4{\pi}G_{\rm eff}a^2}{{\cal H}^2} (\bar{\rho}_m + \bar{p}_m) \nn
&\;  \hspace{1.3cm} + (1+\alpha_B-\alpha_H) \gamma_{\cal H} \Big] \nn
&\; - 24{\pi}G_{\rm eff} a^3 \dfrac{\alpha_B \bar{p}'_m}{\gamma_0 {\cal H}^3} ,
\end{align}
with $G_{\rm eff}$ and $\gamma_{\cal H}$ being as given by \eqref{Friedmann} and \eqref{dPsidtApp}, respectively, and we have
\begin{align}
\gamma_4 \equiv &\; 6 \gamma^{-1}_0\left[ \dfrac{\alpha'_B}{{\cal H}} + (1+\alpha_B) \gamma_{\cal H} + \dfrac{4{\pi}G_{\rm eff}a^2}{{\cal H}^2} (\bar{\rho}_m + \bar{p}_m) \right] \nn
& +\; \dfrac{2\alpha_H k^2}{\gamma_0 {\cal H}^2}, \\
\gamma_5 \equiv &\; -(3+\alpha_M) - \dfrac{\alpha'_K}{\gamma_0{\cal H}} + 6(1-\alpha_B) \dfrac{\alpha'_B}{\gamma_0{\cal H}} \nn
& +\; 2(\alpha_H - \alpha_B) \dfrac{k^2}{\gamma_0 {\cal H}^2} + \dfrac{24{\pi}G_{\rm eff}a^2}{\gamma_0 {\cal H}^2} (\bar{\rho}_m + \bar{p}_m) (1+\alpha_B) \nn
& -\; \Big[6\alpha^2_B + 2\alpha_K - 12\alpha_B -6 \Big] \dfrac{\gamma_{\cal H}}{\gamma_0}, \\
\gamma_6 \equiv &\; \dfrac{2k^2}{\gamma_0 {\cal H}^2} \Big[\alpha_M + \alpha_H(1+\alpha_M) - \alpha_T - \dfrac{\alpha'_H}{{\cal H}} \Big] .
\end{align}
Note that, except in \cite{Duniya:2019mpr}, the parameters $\gamma_1,\, \gamma_2,\, \cdots,\, \gamma_6$ are not the same as those in the literature (e.g.~\cite{Gleyzes:2014rba, Lombriser:2015cla, Sakstein:2016ggl, Gleyzes:2014qga, Gleyzes:2014dya, Bellini:2014fua}).

Moreover, by taking the time derivative of \eqref{dpi-dt}, we use \eqref{dPsidtApp} and \eqref{ddpiddt} to get
\begin{align}\label{dPhidt}
\Phi' + (1+\lambda_1) {\cal H}\Phi =&\; 4{\pi}G_{\rm eff}a^2 \left[ 2\lambda_5 {\cal H} \sigma_m - \lambda_4 q_m - 6\lambda_6 \dfrac{\delta{p}_m}{{\cal H}} \right] \nn
& +\; \lambda_2 {\cal H}\Psi + \lambda_3 {\cal H}^2\Pi, 
\end{align}
where $\lambda_1$, $\lambda_2$, $\lambda_3$, and $\lambda_4$ are as given in \S\ref{sec:UDE}, and
\begin{align}
\lambda_5 \equiv &\; \alpha_M +\dfrac{\alpha'_H}{ {\cal H}\alpha_H} - \dfrac{\sigma'_m}{ {\cal H}\sigma_m} -2\Big(1 + \alpha_H \dfrac{\alpha_B k^2}{\gamma_0 {\cal H}^2}\Big) + \dfrac{\beta_1}{\alpha_H},
\end{align}
with $\lambda_6 \equiv \alpha_H {\alpha_B}/{\gamma_0}$ and, $\beta_1$ being as given in \S\ref{sec:UDE}, and
\begin{align}
\beta_2 \equiv &\; (\alpha_M -\alpha_T)\dfrac{\alpha'_H}{\alpha_H {\cal H}} + (\alpha_H -\alpha_M +\alpha_H\gamma_4 - 1)\dfrac{{\cal H}'}{{\cal H}^2} \nn
& +\; \lambda_4\left[ 1+\alpha_B - \dfrac{4{\pi}G_{\rm eff}a^2}{{\cal H}^2}(\bar{\rho}_m + \bar{p}_m)\right] \nn
& +\; \dfrac{\alpha'_T -\alpha'_M}{{\cal H}} -\alpha_H\gamma_3 .
\end{align}


\section{The Energy-momentum Tensor}\label{App:T_munu} We use the energy-momentum tensor for fluids:
\begin{align}\label{T_munu}
T^0\/_0 =&\; -\left(\bar{\rho} + \delta\rho\right),\quad T^0\/_j = \left(\bar{\rho} + \bar{p}\right)\nabla_j V, \\
T^i\/_j =&\; \left(\bar{p}+\delta{p}\right)\delta^i\/_j + \left(\nabla^i\nabla_j - \dfrac{1}{3}\delta^i\/_j \nabla^2\right) \sigma,
\end{align}
where $\bar{\rho}$ and $\bar{p}$ are the background energy density and pressure, respectively; with $\delta{\rho}$ and $\delta{p}$ being the perturbed energy density and pressure, respectively; $\sigma$ is the anisotropic stress potential, $V$ being the (gauge-invariant) velocity potential.

\bibliographystyle{elsarticle-num} 
\bibliography{magnification_in_beyond_Horndeski}


\end{document}